%% file: RectangularMF_arxiv.tex
\documentclass[english]{article}
\usepackage[T1]{fontenc}
\usepackage[latin9]{inputenc}
\usepackage{geometry}
\usepackage{mathtools}
\usepackage{bm}
\usepackage{amsmath}
\usepackage{amssymb}
\usepackage{mathdots}
\usepackage{graphicx}
\usepackage{microtype}
  
\usepackage{url}\usepackage{algorithmic}\usepackage{bm}
\usepackage{booktabs}\usepackage{comment} 
\usepackage{amsfonts}
\usepackage{color}
\usepackage{algorithm}
\usepackage{subfigure}
\usepackage{latexsym}
\usepackage{psfrag}\usepackage{wrapfig}\usepackage{breqn}\usepackage{booktabs}
\usepackage{xspace}\usepackage{paralist}
\allowdisplaybreaks
\usepackage{amsthm}

\let\hat\widehat
\let\tilde\widetilde

\newcommand{\bx}{\bm{x}}
\newcommand{\by}{\bm{y}}

\newcommand{\bA}{\bm{A}}

\newcommand{\bK}{\bm{K}}

\newcommand{\bM}{\bm{M}}

\newcommand{\bQ}{\bm{Q}}
\newcommand{\bR}{\bm{R}}

\newcommand{\bU}{\bm{U}}
\newcommand{\bV}{\bm{V}}

\newcommand{\bX}{\bm{X}}
\newcommand{\bY}{\bm{Y}}
\newcommand{\bZ}{\bm{Z}}

\newcommand{\F}{{\mathrm{F}}}

\newcommand{\argmin}{\mathop{\mathrm{argmin}}}

\usepackage{enumitem}
\setlist[itemize]{leftmargin=1em}
\setlist[enumerate]{leftmargin=1em}

\definecolor{yxc}{RGB}{255,0,0}
\definecolor{yjc}{RGB}{125,0,0}
\definecolor{cm}{RGB}{0,0,200}
\definecolor{kzw}{RGB}{0,150,0}

\newtheorem{lemma}{\textbf{Lemma}}\newtheorem{theorem}{\textbf{Theorem}}\newtheorem{definition}{\textbf{Definition}}

\title{Beyond Procrustes: Balancing-Free Gradient Descent for Asymmetric Low-Rank Matrix Sensing}

\author{Cong Ma\thanks{C. Ma is with Department of Electrical Engineering and Computer Science, UC Berkeley, Berkeley, CA 94720, USA; Email:
		\texttt{congm@berkeley.edu}.} \\
		 University of California, Berkeley \\
		\and
		 Yuanxin Li\thanks{Y. Li was with Department of Electrical and Computer Engineering, Carnegie Mellon University, Pittsburgh, PA 15213, USA.}  \\
	Carnegie Mellon University\\
	\and
	Yuejie Chi\thanks{Y. Chi is with Department of Electrical and Computer Engineering, Carnegie Mellon University, Pittsburgh, PA 15213, USA; Email: \texttt{yuejiechi@cmu.edu}.} \\
		Carnegie Mellon University }

 \date{}

\begin{document}
\maketitle

\input{abstract.tex}

\input{intro.tex}

\input{results.tex}

\input{analysis.tex}

\input{conclusions.tex}

\appendix
\input{appendix.tex}

\section*{Acknowledgments}
 
This work is supported in part by ONR under the grants N00014-18-1-2142 and N00014-19-1-2404,
by ARO under the grant W911NF-18-1-0303, and by NSF under the grants
CAREER ECCS-1818571, CCF-1806154 and CCF-1901199. A preliminary version of this paper was presented at the 2019 Asilomar Conference on Signals, Systems, and Computers \cite{ma2019beyond}. 

\bibliographystyle{IEEEtran}
\bibliography{bibfileNonconvex_TSP}

\end{document}

%% file: abstract.tex
\begin{abstract}
Low-rank matrix estimation plays a central role in various applications
across science and engineering. Recently, nonconvex formulations based
on matrix factorization are provably solved by simple gradient descent
algorithms with strong computational and statistical guarantees. However,
when the low-rank matrices are asymmetric, existing approaches rely
on adding a regularization term to balance the scale of the two matrix
factors which in practice can be removed safely without hurting the
performance when initialized via the spectral method. In this paper,
we provide a theoretical justification to this   for the matrix sensing problem, which
aims to recover a low-rank matrix from a small number of linear measurements.
As long as the measurement ensemble satisfies the restricted isometry
property, gradient descent---in conjunction with spectral initialization---converges linearly without the need of
explicitly promoting balancedness of the factors; in fact, the factors
stay balanced automatically throughout the execution of the algorithm.
Our analysis is based on analyzing the evolution of a new distance
metric that directly accounts for the ambiguity due to invertible
transforms, and might be of independent interest.

\end{abstract}

\noindent \textbf{Keywords:}
asymmetric low-rank matrix sensing, nonconvex optimization, gradient descent

%% file: intro.tex

\section{Introduction}

Low-rank matrix estimation plays a central role in many applications~\cite{CanTao10,chen2018harnessing,davenport2016overview}.
Broadly speaking, we are interested in estimating a rank-$r$ matrix
$\bM_{\star}=\bX_{\star}\bY_{\star}^{\top}\in\mathbb{R}^{n_{1}\times n_{2}}$
by solving a rank-constrained optimization problem: 
\begin{equation}
\min_{\bm{M}\in\mathbb{R}^{n_{1}\times n_{2}}}\;\mathcal{L}(\bm{M})\quad\mbox{subject to}\quad\mbox{rank}(\bM)\leq r,\label{eq:rank_constrained}
\end{equation}
where $\mathcal{L}(\cdot)$ denotes a certain loss function with the
rank $r$  typically much smaller than the dimension of the matrix.
To reduce computational complexity, a common approach, popularized
by the work of Burer and Monteiro~\cite{burer2003nonlinear,bhojanapalli2016dropping,boumal2016non}, is
to factorize $\bM=\bX\bY^{\top}$ with $\bX\in\mathbb{R}^{n_{1}\times r}$
and $\bY\in\mathbb{R}^{n_{2}\times r}$, and rewrite the above problem~(\ref{eq:rank_constrained})
into an unconstrained nonconvex optimization problem: 
\begin{equation}
\min_{\bm{X}\in\mathbb{R}^{n_{1}\times r},\bm{Y}\in\mathbb{R}^{n_{2}\times r}}\; f(\bX,\bY)\triangleq\mathcal{L}(\bm{X}\bm{Y}^{\top}).\label{eq:general_loss}
\end{equation}
Despite nonconvexity, one might be tempted to estimate the low-rank
factors $(\bX,\bY)$ via gradient descent, which proceeds via the
following update rule 
\begin{align}
\begin{bmatrix}\bm{X}_{t+1}\\
\bm{Y}_{t+1}
\end{bmatrix} & =\begin{bmatrix}\bm{X}_{t}\\
\bm{Y}_{t}
\end{bmatrix}-\eta_{t}\begin{bmatrix}\nabla_{\bX}f(\bm{X}_{t},\bm{Y}_{t})\\
\nabla_{\bY}f(\bm{X}_{t},\bm{Y}_{t})
\end{bmatrix} \label{eq:vgd}
\end{align}
from $(\bX_{0},\bY_{0})$ some
proper initialization. Here, $\eta_{t}$ is the step size, $\nabla_{\bX}f$ and $\nabla_{\bY}f$ are the gradients of $f$ w.r.t. $\bX$ and $\bY$, respectively.

Significant progress has been made recently in understanding the performance
of gradient descent for nonconvex matrix estimation. Somewhat surprisingly,
most of the existing guarantees are not directly applicable to the
vanilla gradient descent rule \eqref{eq:vgd}. One particular challenge
is associated with the identifiability of the factors $(\bm{X},\bm{Y})$---they are indistinguishable as long as their product $\bm{X}\bm{Y}^{\top}$
is the same. What is worse, if the norms of the factors become highly
unbalanced, gradient descent might diverge easily. Consequently, it
becomes a routine procedure to insert a regularizer~$g(\bm{X},\bm{Y})$
that balances the two factors \cite{tu2015low,zheng2016convergence,park2018finding}:
\begin{equation}
g(\bX,\bY)\triangleq\lambda\|\bX^{\top}\bX-\bY^{\top}\bY\|_{\F}^{2},\label{eq:balancing_term}
\end{equation}
where $\lambda>0$ is some regularization parameter, and apply gradient
descent to the regularized loss function instead: 
\begin{equation}
\min_{\bm{X}\in\mathbb{R}^{n_{1}\times r},\bm{Y}\in\mathbb{R}^{n_{2}\times r}}  f_{\mathrm{reg}}(\bX,\bY)\triangleq f(\bX,\bY)+g(\bX,\bY).\label{eq:reg_loss}
\end{equation}
For a variety of important problems such as low-rank matrix sensing
and matrix completion, it has been established that gradient descent
over the regularized loss function, when properly initialized, achieves
compelling statistical and computational guarantees.

\subsection{Why balancing is needed in prior work?}


Before we investigate the possibility of a balancing-free procedure (i.e.~vanilla gradient descent as in~(\ref{eq:vgd})), let us first explain using a heuristic argument why balancing is needed in the prior literature. 

To handle the asymmetric factorization, it is common to stack the
two factors into one augmented factor $\bZ_{\star}\triangleq\begin{bmatrix}\bX_{\star}\\
\bY_{\star}
\end{bmatrix}\in\mathbb{R}^{(n_{1}+n_{2})\times r}$ and then seek to estimate $\bZ_{\star}$ directly, by rewriting the
loss function with respect to the lifted low-rank matrix: $\bZ_{\star}\bZ_{\star}^{\top}=\begin{bmatrix}\bX_{\star}\bX_{\star}^{\top} & \bX_{\star}\bY_{\star}^{\top}\\
\bY_{\star}\bX_{\star}^{\top} & \bY_{\star}\bY_{\star}^{\top}
\end{bmatrix}\in\mathbb{R}^{(n_{1}+n_{2})\times(n_{1}+n_{2})}$. It is obvious that the loss function originally with respect to
the asymmetric matrix $\bX_{\star}\bY_{\star}^{\top}$ only constrains
the off-diagonal blocks of $\bZ_{\star}\bZ_{\star}^{\top}$ and not
the diagonal ones; correspondingly, the loss function is not (restricted)
strongly convex with respect to the augmented factor, unless we appropriately
regularize the diagonal blocks. This gives rise to the adoption of
the regularization term in \eqref{eq:balancing_term}.

\begin{figure}[t]
\begin{centering}
\begin{tabular}{cc}
 \includegraphics[width=0.42\textwidth]{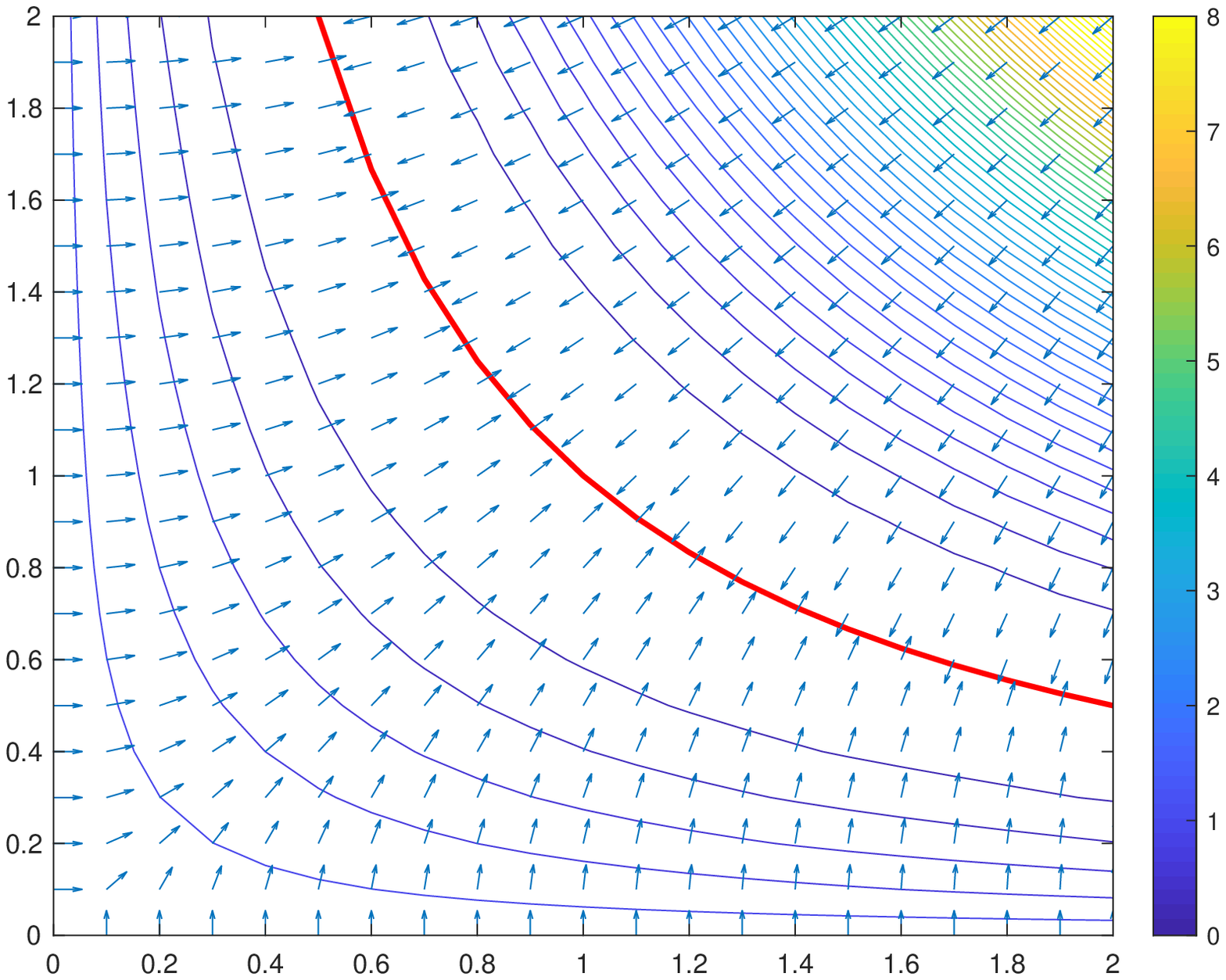}  & \includegraphics[width=0.42\textwidth]{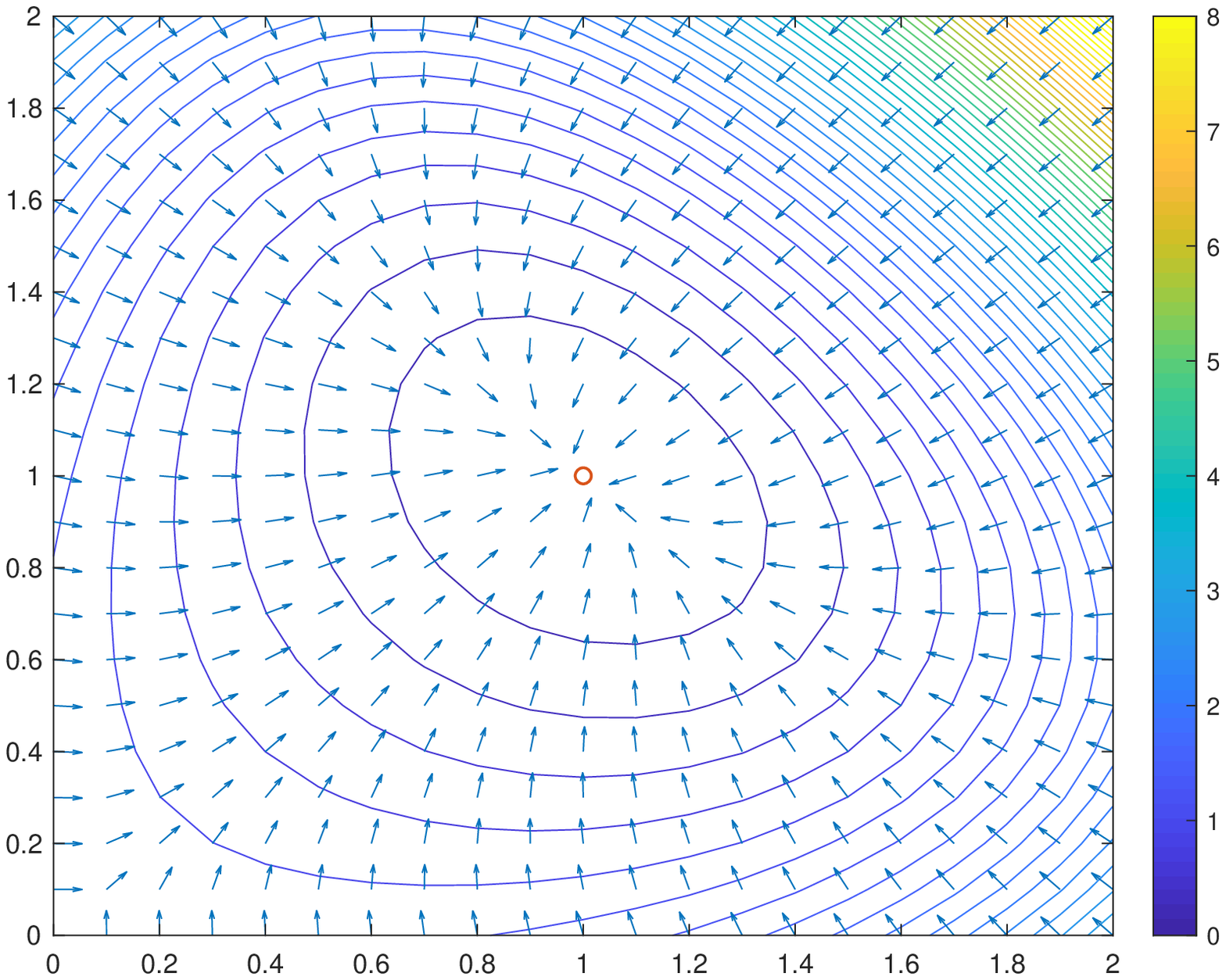} \tabularnewline
(a) unregularized loss $f$  & (b) regularized loss $f_{\mathrm{reg}}$ \tabularnewline
\end{tabular}
\par\end{centering}
\caption{The geometry for the scalar case $f(x,y)=(xy-1)^{2}$ and $g(x,y)=(x^{2}-y^{2})^2/8$.
The regularized loss function is locally strongly convex while the
unregularized one is nonconvex; in particular, the Hessian of the
unregularized loss function is rank deficient on the ambiguity set
$xy=1$ (colored in red). }
\label{fig:landscape_illustration} 
\end{figure}
To develop more intuitions regarding why this regularization term
\eqref{eq:balancing_term} may help analysis, consider a toy example
of factorizing a rank-one matrix $\bm{x}_{\star}\bm{y}_{\star}^{\top}$,
where $f(\bm{x},\bm{y})$ and $g(\bm{x},\bm{y})$ respectively are
$f(\bm{x},\bm{y})=\|\bm{x}\bm{y}^{\top}-\bm{x}_{\star}\bm{y}_{\star}^{\top}\|_{\F}^{2}/2$
and $g(\bm{x},\bm{y})=(\|\bm{x}\|_{2}^{2}-\|\bm{y}\|_{2}^{2})^{2}/8$.
Figure~\ref{fig:landscape_illustration} illustrates the landscape
of the unregularized loss function $f({x},{y})$ and the regularized
loss function $f_{\mathrm{reg}}({x},{y})$, respectively, when
the arguments are scalar-valued, i.e.~$n_{1}=n_{2}=1$. One can clearly
appreciate the value of the regularizer: $f_{\mathrm{reg}}({x},{y})$
becomes strongly convex in the local neighborhood around the global
optimum $(1,1)$. In contrast, the Hessian of the unregularized loss
function $f_{\mathrm{reg}}({x},{y})$ remains rank deficient along
the ambiguity set $\{(x,y)\,|\,xy=1\}$, making the analysis less
tractable.

\subsection{This paper: balancing-free procedure?}

\begin{figure}[t]
\centering \includegraphics[width=0.42\textwidth]{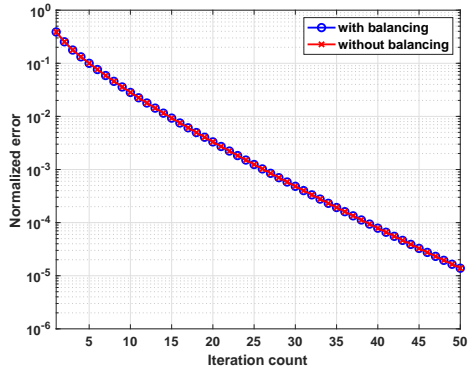}
\caption{The normalized reconstruction error $\|\bX_{t}\bY_{t}^{\top}-\bM_{\star}\|_{\mathrm{F}}/\|\bM_{\star}\|_{\mathrm{F}}$
with respect to the iteration count, for completing a $1000\times1000$
matrix of rank-$10$ when each entry is observed independently with probability
$p=0.15$. The balancing regularizer is set as $g(\bX,\bY)=\frac{1}{64}\|\bX^{\top}\bX-\bY^{\top}\bY\|_{\mathrm{F}}^{2}$
following the suggestion in \cite{yi2016fast}.}
\label{fig:rectangularMC} 
\end{figure}

The goal of this paper is to understand the effectiveness of vanilla
gradient descent \eqref{eq:vgd} when initialized with balanced factors.
Indeed, Figure~\ref{fig:rectangularMC} plots the normalized error
$\|\bX_{t}\bY_{t}^{\top}-\bM_{\star}\|_{\mathrm{F}}/\|\bM_{\star}\|_{\mathrm{F}}$
for low-rank matrix completion, which aims to recover a low-rank matrix from a subset of its observations \cite{chi2018low}, with
respect to the iteration count, using either a regularized loss function
or an unregularized loss function when initialized by the spectral
method. The two sequences of iterates converge in almost exactly the same trajectory,
suggesting that gradient descent over the unregularized loss function
converges almost in the same manner as its regularized counterpart,
and perhaps is more natural to use in practice since it eliminates
the tuning of the regularization parameters.

This paper justifies formally that even without explicit balancing
in asymmetric low-rank matrix sensing, gradient descent converges linearly
towards the global optimum, as long as the initialization is (nearly)
balanced and close to the optimum. As will be detailed later, our analysis is simple and
built on a novel distance metric that directly accounts for the ambiguity
due to invertible transformations---in contrast, the ambiguity set
reduces to orthonormal transforms when the balancing regularization
is present. Our key message is this:

{ \setlist{rightmargin=\leftmargin} 
\begin{itemize}
\item[] \emph{As long as the factors are (nearly) balanced in a basin of attraction at the initialization,
they will stay approximately balanced throughout the trajectory of
gradient descent, and therefore no additional regularization is necessary.
} 
\end{itemize}

\subsection{Notation}

We use boldface lowercase (resp.~uppercase) letters to represent
vectors (resp.~matrices). We denote by $\left\Vert \bx\right\Vert _{2}$
the $\ell_{2}$ norm of a vector $\bx$, and $\bX^{\top}$, $\bX^{-1}$,
$\left\Vert \bX\right\Vert $ and $\left\Vert \bX\right\Vert _{\F}$
the transpose, the inverse, the spectral norm and the Frobenius norm
of a matrix $\bX$, respectively. Furthermore, we denote $\bX^{-\top}=(\bX^{-1})^{\top}=(\bX^{\top})^{-1}$
for an invertible matrix $\bX$. The $k$th largest singular value
of a matrix $\bX$ is denoted by $\sigma_{k}(\bX)$. The inner product
between two matrices $\bX$ and $\bY$ is defined as $\langle\bX,\bY\rangle=\mathrm{Tr}\left(\bY^{\top}\bX\right)$,
where $\mathrm{Tr}\left(\cdot\right)$ denotes the trace operator.
Denote by $\mathcal{O}^{r\times r}$ the set of $r\times r$ orthonormal
matrices. In addition, we use $c$ and $C$ with different subscripts
to represent positive numerical constants, whose values may change
from line to line. }

%% file: results.tex
\section{Main Results}


Let the object of interest $\bm{M}_{\star}\in\mathbb{R}^{n_{1}\times n_{2}}$
be a rank-$r$ matrix whose compact Singular Value Decomposition (SVD)
is given by 
\[
\bm{M}_{\star}=\bm{U}_{\star}\bm{\Sigma}_{\star}\bm{V}_{\star}^{\top},
\]
where $\bm{U}_{\star}\in\mathbb{R}^{n_{1}\times r}$, $\bm{V}_{\star}\in\mathbb{R}^{n_{2}\times r}$
and $\bm{\Sigma}_{\star}\in\mathbb{R}^{r\times r}$ correspond to
the left singular vectors, the right singular vectors and the singular
values, respectively. Without loss of generality, we denote the ground
truth factors as 
\begin{equation}
\bm{X}_{\star}\triangleq\bm{U}_{\star}\bm{\Sigma}_{\star}^{1/2},\qquad\text{and}\qquad\bm{Y}_{\star}\triangleq\bm{V}_{\star}\bm{\Sigma}_{\star}^{1/2}.\label{eq:ground_truth}
\end{equation}
Let $\sigma_{\max}\triangleq\sigma_{1}(\bm{M}_{\star})$ (resp.~$\sigma_{\min}\triangleq\sigma_{r}(\bm{M}_{\star})$)
be the largest (resp.~smallest) nonzero singular value of $\bm{M}_{\star}$. 
The condition number of $\bm{M}_{\star}$ is therefore defined as
$\kappa\triangleq\sigma_{\max}/\sigma_{\min}$. 

Since the factors are identifiable up to invertible transforms, i.e.~$(\bm{X}_{\star}\bm{P})(\bm{Y}_{\star}\bm{P}^{-\top})^{\top}=\bm{X}_{\star}\bm{Y}_{\star}^{\top}$
for any invertible matrix $\bm{P}\in\mathbb{R}^{r\times r}$, it is natural to measure
the distance between two pairs of factors $\bm{Z}=\left[\begin{array}{c}
\bm{X}\\
\bm{Y}
\end{array}\right]\in\mathbb{R}^{(n_{1}+n_{2})\times r}$ and $\bm{Z}_{\star}=\left[\begin{array}{c}
\bm{X}_{\star}\\
\bm{Y}_{\star}
\end{array}\right]\in\mathbb{R}^{(n_{1}+n_{2})\times r}$ via the following function:\footnote{More rigorously, we should write $\inf$ instead of $\min$ in the
definition of $\mathrm{dist}(\cdot,\cdot)$. However, as we will soon
see, in the cases we care about, the minimum can always be achieved by some
invertible matrix $\bm{P}$.} 
\begin{equation*}
\mathrm{dist}\left(\bm{Z},\bm{Z}_{\star}\right)=\min_{\substack{\bm{P}\in\mathbb{R}^{r\times r} \\ \text{invertible}}}\sqrt{\left\Vert \bm{X}\bm{P}-\bm{X}_{\star}\right\Vert _{\mathrm{F}}^{2}+\left\Vert \bm{Y}\bm{P}^{-\top}-\bm{Y}_{\star}\right\Vert _{\mathrm{F}}^{2}}. 
\end{equation*}

\subsection{Low-rank matrix sensing}

Low-rank matrix sensing refers to the problem of recovering a low-rank matrix (i.e.~$\bm{M}_\star$) from a small number of linear measurements. Specifically, we are given a set of $m$ measurements as follows 
\begin{equation}
y_{i}=\langle\bA_{i},\bM_{\star}\rangle=\langle\bA_{i},\bX_{\star}\bY_{\star}^{\top}\rangle,\qquad i=1,\cdots,m,\label{eq:sensing}
\end{equation}
where $\bA_{i}\in\mathbb{R}^{n_{1}\times n_{2}}$ is the $i$th sensing
matrix. For convenience, we define $\mathcal{A}:\mathbb{R}^{n_{1}\times n_{2}}\to\mathbb{R}^{m}$
as an affine transformation from $\mathbb{R}^{n_{1}\times n_{2}}$
to $\mathbb{R}^{m}$, such that $\mathcal{A}\left(\bM\right)=\left\{ \langle\bA_{i},\bM\rangle\right\} _{1\leq i\leq m}$.
Consequently, one can compactly write (\ref{eq:sensing}) as $\by=\mathcal{A}\left(\bM_{\star}\right)$.
The adjoint operator $\mathcal{A}^{*}:\mathbb{R}^{m}\to\mathbb{R}^{n_{1}\times n_{2}}$
is defined as $\mathcal{A}^{*}(\bm{y})=\sum_{i=1}^{m}y_{i}\bm{A}_{i}$.

To recover the low-rank matrix, a natural choice is to minimize the
least-squares loss function 
\begin{equation}
f\left(\bX,\bY\right)\triangleq\frac{1}{2}\left\Vert \bm{y}-\mathcal{A}(\bX\bY^{\top})\right\Vert _{2}^{2}.\label{eq:squared_loss_sensing}
\end{equation}
Algorithm~\ref{alg:matrix_sensing} describes the gradient descent
algorithm initialized by the spectral method \cite{chen2020spectral} for minimizing~\eqref{eq:squared_loss_sensing}.
Compared to the Procrustes Flow (PF) algorithm in \cite{tu2015low},
which minimizes the regularized loss function in \eqref{eq:reg_loss},
the new algorithm does not include the balancing regularizer $g(\bX,\bY)$.


\begin{algorithm}[t]
\caption{Gradient Descent with Spectral Initialization (unregularized Procrustes
Flow)}
\label{alg:matrix_sensing} \textbf{Input:} Measurements $\by=\left\{ y_{i}\right\} _{1\leq i\leq m}$,
and sensing matrices $\left\{ \bA_{i}\right\} _{1\leq i\leq m}$.

\noindent \textbf{Parameters:} Step size $\eta_{t}$, rank $r$, and
number of iterations $T$.

\noindent \textbf{Initialization:} Initialize $\bX_{0}=\bU\boldsymbol{\Sigma}^{1/2}$
and $\bY_{0}=\bV\boldsymbol{\Sigma}^{1/2}$, where $\bU\boldsymbol{\Sigma}\bV^{\top}$
is the rank-$r$ SVD of the surrogate matrix $\bK=\frac{1}{m}\mathcal{A}^{*}(\bm{y})=\frac{1}{m}\sum_{i=1}^{m}y_{i}\bA_{i}$.

\noindent \textbf{Gradient loop:} For $t=0,1,\cdots,T-1$, do \begin{subequations}\label{subeq:gd_sensing}
\begin{align}
\bX_{t+1} & =\bX_{t}-\frac{\eta_{t}}{\left\Vert \bY_{0}\right\Vert ^{2}}\cdot\left[\sum_{i=1}^{m}\left(\langle\bA_{i},\bX_{t}\bY_{t}^{\top}\rangle-y_{i}\right)\bA_{i}\bY_{t}\right];\\
\bY_{t+1} & =\bY_{t}-\frac{\eta_{t}}{\left\Vert \bX_{0}\right\Vert ^{2}}\cdot\left[\sum_{i=1}^{m}\left(\langle\bA_{i},\bX_{t}\bY_{t}^{\top}\rangle-y_{i}\right)\bA_{i}^{\top}\bX_{t}\right].
\end{align}
\end{subequations}

\noindent \textbf{Output:} $\bX_{T}$ and $\bY_{T}$. 
\end{algorithm}

\subsection{Theoretical guarantee for local linear convergence}

To understand the performance of Algorithm~\ref{alg:matrix_sensing},
we adopt a standard assumption on the sensing operator $\mathcal{A}$,
namely the Restricted Isometry Property (RIP). \begin{definition}[RIP]
The operator $\mathcal{A}(\cdot)$ is said to satisfy the rank-$r$
RIP with a constant $\delta_{r}\in[0,1)$, if 
\[
(1-\delta_{r})\left\Vert \bM\right\Vert _{\F}^{2}\le\left\Vert \mathcal{A}(\bM)\right\Vert _{2}^{2}\le(1+\delta_{r})\left\Vert \bM\right\Vert _{\F}^{2}
\]
holds for all matrices $\bM\in\mathbb{R}^{n_{1}\times n_{2}}$ of
rank at most $r$. \end{definition}

It is well-known that many measurement ensembles satisfy the RIP property
\cite{recht2010guaranteed}. For example, under the Gaussian design where the entries of $\bA_{i}$'s
are composed of i.i.d.~Gaussian entries $\mathcal{N}(0,1/m)$, 
the RIP is satisfied as long as $m$ is on the order of $(n_{1}+n_{2})r/\delta_{r}^{2}$.

Armed with the RIP, we have the following theoretical guarantee for the
local convergence of Algorithm~\ref{alg:matrix_sensing}. 

\begin{theorem}\label{thm:matrix_sensing} Suppose that $\mathcal{A}(\cdot)$
satisfies the RIP with $\delta_{2r}\leq c$ for some sufficiently
small constant $c$. Let $\bm{Z}_{0}\triangleq\left[\begin{array}{c}
\bm{X}_{0}\\
\bm{Y}_{0}
\end{array}\right]$ be any initialization point that satisfies 
\begin{equation}
\min_{\bm{R}\in\mathcal{O}^{r\times r}}\left\Vert \bm{Z}_{0}\bm{R}-\bm{Z}_{\star}\right\Vert _{\mathrm{F}}\leq c_{0}\frac{1}{\kappa^{3/2}}\sigma_{\min}^{1/2} ,\label{eq:init_basin}
\end{equation}
for some small enough constant $c_{0}>0$. Then there exist some constant
$c_{1}>0$ such that at as long as $\eta_{t}=\eta=c_{1}$, the iterates
of unregularized gradient descent (cf.~(\ref{subeq:gd_sensing})) satisfy 
\[
\mathrm{dist}\left(\bm{Z}_{t},\bm{Z}_{\star}\right)\leq\left(1-\frac{\eta}{50\kappa}\right)^{t}\mathrm{dist}\left(\bm{Z}_{0},\bm{Z}_{\star}\right).
\]
\end{theorem}

In words, Theorem~\ref{thm:matrix_sensing} reveals that if the initialization
$\bZ_{0}$ lands in a basin of attraction given by \eqref{eq:init_basin},
then Algorithm~\ref{alg:matrix_sensing} converges linearly with
a constant step size. To reach $\epsilon$-accuracy, i.e.~$\mathrm{dist}\left(\bm{Z}_{t},\bm{Z}_{\star}\right)\leq\epsilon$,
it takes an order of $\kappa\log(1/\epsilon)$ iterations, which is
order-wise equivalent to the regularized PF algorithm proposed in~\cite{tu2015low}.
Comparing to \cite{tu2015low}, which requires $\delta_{6r}\leq c$,
Theorem~\ref{thm:matrix_sensing} only requires a weaker assumption
$\delta_{2r}\leq c$. However, the basin of attraction allowed by
Theorem~\ref{thm:matrix_sensing} is smaller than that in \cite{tu2015low},
which is specified by $\min_{\bm{R}\in\mathcal{O}^{r\times r}}\|\bm{Z}_{0}\bm{R}-\bm{Z}_{\star}\|_{\mathrm{F}}\leq c_{0} \sigma_{\min}^{1/2}$. Compared with prior work that relies on local strong convexity to establish linear convergence, our result  suggests the benign behavior of gradient descent even
in the absence of local strong convexity.

\subsection{Achieving global convergence with a proper initialization}
We are still in need of finding a good initialization that obeys \eqref{eq:init_basin}.
In general, one could initialize with the balanced factors of the
output of projected gradient descent
(over the low-rank matrix), i.e. 
\[
\bm{M}_{\tau+1}=\mathcal{P}_{r}\left(\bm{M}_{\tau}-\frac{1}{m}\sum_{i=1}^{m}\left(\langle\bA_{i},\bm{M}_{\tau}\rangle-y_{i}\right)\bA_{i}\right),
\]
where $\mathcal{P}_{r}(\cdot)$ is the Euclidean projection operator
to the set of rank-$r$ matrices. The spectral initialization specified
in Algorithm~\ref{alg:matrix_sensing} can be regarded as the output
at the first iteration, initialized at zero $\bm{M}_{0}=\bm{0}$.
Based on \cite{oymak2018sharp,tu2015low}, the balanced factorization of $\bm{M}_{\tau}$, denoted by $\widetilde{\bm{Z}}_{\tau}$, satisfy 
\begin{equation}
\min_{\bR\in\mathcal{O}^{r\times r}}\big\Vert \widetilde{\bZ}_{\tau}\bR-\bZ_{\star}\big\Vert _{\F}\le c_{2}(2\delta_{4r})^{\tau}\frac{\left\Vert \bM_{\star}\right\Vert _{\F}}{\sigma_{\min}^{1/2}}\label{eq:PGD}
\end{equation}
for some constant $c_{2}$. Thus, to achieve the required initialization
condition~(\ref{eq:init_basin}) using the spectral method specified
in Algorithm~\ref{alg:matrix_sensing} (which corresponds to setting $\bm{Z}_0 = \widetilde{\bm{Z}}_{1}$ with
$\tau=1$ in \eqref{eq:PGD}), we need 
\[
\delta_{4r}\le c_{2}\frac{1}{\kappa^{3/2}}\cdot\frac{\sigma_{\min}}{\left\Vert \bM_{\star}\right\Vert _{\F}}.
\]
In particular, under Gaussian design, where each measurement matrix $\bm{A}_{i}$ has i.i.d.~$\mathcal{N}(0,1/m)$ entries, a total of $m \gtrsim nr  \kappa^3 \left\Vert \bM_{\star}\right\Vert _{\F}^2 / \sigma_{\min}^2$ measurements suffice for the above requirement on $\delta_{4r}$. This is worse by a factor of  $\kappa^3$ compared with the sample complexity guarantee in \cite{tu2015low} with the balancing regularizer, which is due to the restriction on the basin of attraction, as we have remarked earlier. Improving the dependence on $\kappa$ is an interesting future direction.

In order to alleviate the dependency on the condition number $\kappa$, we can allow a few iterations of \eqref{eq:PGD} and set the initialization as $\bm{Z}_0   = \widetilde{\bm{Z}}_{\tau}$, a procedure suggested by \cite{tu2015low}. The advantage of this hybrid procedure is that the switch to factored gradient descent allows a smaller per-iteration memory and computation complexity after the iterates of projected gradient descent enter the basin of attraction. Consequently, the algorithm is still guaranteed to  succeed when $\delta_{4r}\leq\delta_{c}$ 
for a sufficiently small constant $\delta_{c}$ (which implies a near-optimal sample complexity of $m\gtrsim nr$ under Gaussian design), by running at least
\[
\tau\ge c_{1}\log{\left(\kappa^{3/2}\frac{\left\Vert \bM_{\star}\right\Vert _{\F}}{\sigma_{\min}}\right)}/\log{\left(\delta_{c}^{-1}\right)}
\]
iterations of projected gradient descent for initialization, which order-wise
matches the requirement in \cite{tu2015low}.

\section{Related Work}

Low-rank matrix estimation has been extensively studied in recent
years \cite{chen2018harnessing,davenport2016overview}, due to its
broad applicability in collaborative filtering, imaging science, and
machine learning, to name a few. Convex relaxation approaches based
on nuclear norm minimization are among the first set of algorithms
with near-optimal statistical guarantees, e. g. \cite{CanTao10,candes2009exact,CanPla10,Negahban2012restricted,Gross2011recovering,Recht2009SimplerMC,chen2013robustSpectralMC,recht2010guaranteed,negahban2011estimation},
however, their computational costs are often prohibitive in practice.

To cope with the computational challenges, a popular approach in practice
is to invoke low-rank matrix factorization and then apply first-order methods
such as gradient descent directly over the factors to recover the
underlying low-rank structure. This approach is demonstrated to possess
near-optimal statistical and computational guarantees in a variety
of low-rank matrix recovery problems, including but not limited to
\cite{tu2015low,zheng2015convergent,keshavan2010few,sun2015guaranteed,jain2013low,ma2017implicit,chen2015fast,li2018nonconvex,li2016deconvolution,chen2018gradient}.
The readers are referred to the recent overview \cite{chi2018nonconvex}
for additional references.

To the best of our knowledge, the balancing regularization term \eqref{eq:balancing_term}
was first introduced in \cite{tu2015low} to deal with asymmetric
matrix factorization, and has become a standard approach to deal with
asymmetric low-rank matrix estimation \cite{zheng2016convergence,park2018finding,yi2016fast,li2017nonconvex,zhang2018fast,chen2019nonconvex}.
A major benefit of adding the regularization term is to reduce the
ambiguity set from invertible transforms to orthonormal transforms. For the special rank-one matrix
recovery problem, there are some evidence in the prior literature
that a balancing regularization is not needed, for example, Ma et~al.
\cite{ma2017implicit} established that vanilla gradient descent works
for blind deconvolution at a near-optimal sample complexity with spectral
initialization. In \cite{du2018algorithmic}, the trajectory of gradient
descent is studied for asymmetric matrix factorization with an infinitesimal
and diminishing step size; in contrast, we consider the case when
the step size is constant for low-rank matrix estimation with incomplete
observations. Finally, very recently, \cite{charisopoulos2019low} also studied low-rank matrix sensing using a nonsmooth formulation without the balancing regularization via subgradient descent. 

Complementary to the algorithmic analysis, we remark that a similar regularization term \eqref{eq:balancing_term}
is also adopted when analyzing the optimization landscape of low-rank
matrix estimation, e.g.~\cite{ge2016matrix,ge2017no,zhu2017global,zhu2018global,li2019symmetry}.
 It is worth mentioning that when converting a nuclear-norm regularized problem into a nonconvex formulation, \cite{li2018non} demonstrated that the nonconvex problem has benign geometry without adding the balancing regularization, since the nuclear norm regularization induces a term $\tfrac{1}{2} (\| \bX\|_{\F}^2 + \| \bY\|_{\F}^2)$ which ensures both factors have similar sizes. Very recently, \cite{li2020global} showed that the balancing regularizer is unnecessary from the landscape analysis perspective.

After the initial version of the current paper, several other works have further examined the balancing-free  low-rank matrix optimization problem. In particular,  Tian, Ma and Chi developed a scaled gradient descent algorithm \cite{tong2020accelerating} that achieves a faster convergence rate independent of the condition number $\kappa$ without imposing the balancing regularization for a variety of low-rank matrix estimation problems, which are further extended in \cite{tong2020low} to achieve robustness to adversarial outliers.

%% file: analysis.tex
\section{Proof of Theorem~\ref{thm:matrix_sensing}}

In this section, we provide the proof of Theorem~\ref{thm:matrix_sensing}.
We first discuss some basic properties of aligning two low-rank factors
via an invertible transformation. Then we prove a similar result for
a warm-up case of low-rank matrix factorization. In the end, viewing
matrix sensing as a perturbed version of low-rank matrix factorization
helps us finish the proof of Theorem~\ref{thm:matrix_sensing}.

\subsection{Alignment via invertible transformations}
We begin with introducing the alignment matrix will play a key role in the subsequent analysis.
\begin{definition}
Fix a matrix $\bm{Z}=\left[\begin{array}{c}
\bm{X}\\
\bm{Y}
\end{array}\right]\in\mathbb{R}^{(n_{1}+n_{2})\times r}$.  We define the optimal alignment matrix $\bm{Q}$ between $\bm{Z}$
and $\bm{Z}_{\star}$ as 
\[
\bm{Q}\triangleq\argmin_{\substack{\bm{P}\in\mathbb{R}^{r\times r} \\ \text{invertible}}}\left\Vert \bm{X}\bm{P}-\bm{X}_{\star}\right\Vert _{\mathrm{F}}^{2}+\left\Vert \bm{Y}\bm{P}^{-\top}-\bm{Y}_{\star}\right\Vert _{\mathrm{F}}^{2},
\]
whenever the minimum is attained. \end{definition}
As we will soon see, for the iterates
$\{\bm{Z}_{t}\}_{t\geq0}$ generated by Algorithm~\ref{alg:matrix_sensing},
the optimal alignment matrix is always well-defined. Furthermore,
we call $\bm{Z}$ and $\bm{Z}_{\star}$ aligned if the corresponding
optimal alignment matrix is just the identity matrix $\bm{I}_{r}$.
Below we provide some basic understandings of this alignment matrix.

The following lemma provides a sufficient condition for the existence
of the optimal alignment matrix.

\begin{lemma}\label{lemma:alignment-near-rotation}Fix some matrix
$\bm{Z}=\left[\begin{array}{c}
\bm{X}\\
\bm{Y}
\end{array}\right]\in\mathbb{R}^{(n_{1}+n_{2})\times r}$. Suppose that there exists a matrix $\bm{P}\in\mathbb{R}^{r\times r}$
with $1/2\leq\sigma_{r}(\bm{P})\leq\sigma_{1}(\bm{P})\leq3/2$ such
that 
\begin{equation}
\max\left\{ \left\Vert \bm{X}\bm{P}-\bm{X}_{\star}\right\Vert _{\mathrm{F}},\left\Vert \bm{Y}\bm{P}^{-\top}-\bm{Y}_{\star}\right\Vert _{\mathrm{F}}\right\} \leq\delta\leq\frac{\sigma_{r}\left(\bm{X}_{\star}\right)}{80}.\label{eq:assumption_alignment_near_rotation}
\end{equation}
Then the optimal alignment matrix $\bm{Q}\in\mathbb{R}^{r\times r}$
between $\bm{Z}$ and $\bm{Z}_{\star}$ exists. In addition, the matrix
$\bm{Q}$ satisfies 
\[
\left\Vert \bm{P}-\bm{Q}\right\Vert \leq\left\Vert \bm{P}-\bm{Q}\right\Vert _{\mathrm{F}}\leq\frac{5\delta}{\sigma_{r}\left(\bm{X}_{\star}\right)}.
\]
\end{lemma}

Next, the lemma below presents a necessary condition for $\bm{Q}$
to be the optimal alignment matrix between $\bm{Z}$ and $\bm{Z}_{\star}$.
\begin{lemma}\label{lemma:alignment}Let $\bm{Z}$ and $\bm{Z}_{\star}$
be any two matrices. Suppose that the optimal alignment matrix $\bm{Q}$
between $\bm{Z}$ and $\bm{Z}_{\star}$ exists. Then we have 
\[
\tilde{\bm{X}}^{\top}(\tilde{\bm{X}}-\bm{X}_{\star})=(\tilde{\bm{Y}}-\bm{Y}_{\star})^{\top}\tilde{\bm{Y}},
\]
where $\tilde{\bm{X}}=\bm{X}\bm{\bm{Q}}$ and $\tilde{\bm{Y}}=\bm{\bm{Y}}\bm{\bm{Q}}^{-\top}$
are two matrices after the alignment. \end{lemma}

Both lemmas provide basic understandings of the solution to the alignment
problem with invertible transformations, which can be regarded as
a generalization of the classical orthogonal Procrustes problem that
only considers orthonormal transformations. Clearly, this generalized
problem is more involved and our work provides some basic
understandings. 

\subsection{A warm-up: low-rank matrix factorization}

We consider the following minimization problem for low-rank matrix
factorization 
\begin{equation}
f_{\mathrm{MF}}\left(\bm{X},\bm{Y}\right)=\frac{1}{2}\left\Vert \bm{X}\bm{Y}^{\top}-\bm{M}_{\star}\right\Vert _{\mathrm{F}}^{2},\label{eq:loss-mf}
\end{equation}
where $\bm{X}\in\mathbb{R}^{n_{1}\times r}$ and $\bm{Y}\in\mathbb{R}^{n_{2}\times r}$.
The gradient descent updates with an initialization $(\bm{X}_{0},\bm{Y}_{0})$
can be written as 
\begin{equation}
\begin{aligned}\bm{X}_{t+1} & =\bm{X}_{t}-\frac{\eta}{\sigma_{\mathrm{\max}}}\nabla_{\bm{X}}f_{\mathrm{MF}}\left(\bm{X}_{t},\bm{Y}_{t}\right) =\bm{X}_{t}-\frac{\eta}{\sigma_{\mathrm{\max}}}(\bm{X}_{t}\bm{Y}_{t}^{\top}-\bm{M}_{\star})\bm{Y}_{t};\\
\bm{Y}_{t+1} & =\bm{Y}_{t}-\frac{\eta}{\sigma_{\mathrm{max}}}\nabla_{\bm{Y}}f_{\mathrm{MF}}\left(\bm{X}_{t},\bm{Y}_{t}\right) =\bm{Y}_{t}-\frac{\eta}{\sigma_{\mathrm{max}}}(\bm{X}_{t}\bm{Y}_{t}^{\top}-\bm{M}_{\star})^{\top}\bm{X}_{t}.
\end{aligned}
\label{eq:gd-update}
\end{equation}
Here, $\eta>0$ stands for the step size. We have the following theorem
regarding the performance of \eqref{eq:gd-update}, which parallels
Theorem~\ref{thm:matrix_sensing}. \begin{theorem}\label{thm:matrix_factorization}
Let $\bm{Z}_{0}=\left[\begin{array}{c}
\bm{X}_{0}\\
\bm{Y}_{0}
\end{array}\right]\in\mathbb{R}^{(n_{1}+n_{2})\times r}$ be any initialization point that satisfies 
\begin{equation}
\min_{\bm{R}\in\mathcal{O}^{r\times r}}\left\Vert \bm{Z}_{0}\bm{R}-\bm{Z}_{\star}\right\Vert _{\mathrm{F}}\leq c_{0}\frac{1}{\kappa^{3/2}} \sigma_{\min}^{1/2}\label{eq:init-MF}
\end{equation}
for some sufficiently small constant $c_{0}>0$. Then setting the
step size $\eta>0$ to be some sufficiently small constant, the iterates
of GD (cf.~(\ref{eq:gd-update})) satisfy 
\begin{align*}
\mathrm{dist}\left(\bm{Z}_{t},\bm{Z}_{\star}\right) & \leq\left(1-\frac{\eta}{50\kappa}\right)^{t}\mathrm{dist}\left(\bm{Z}_{0},\bm{Z}_{\star}\right).
\end{align*}
\end{theorem}

To prove Theorem~\ref{thm:matrix_factorization}, we need the following
properties regarding the gradients of $f_{\mathrm{MF}}(\bX,\bY)$; the proofs
are deferred to the appendix. \begin{lemma}[Gradient dominance]\label{lemma:gradient-dominance}
Suppose that $\bm{Z}=\left[\begin{array}{c}
\bm{X}\\
\bm{Y}
\end{array}\right]\in\mathbb{R}^{(n_{1}+n_{2})\times r}$ is aligned with $\bm{Z}_{\star}$, i.e. 
\[
\bm{I}_{r}=\argmin_{\substack{\bm{P}\in\mathbb{R}^{r\times r} \\ \text{invertible}}} \; \left\Vert \bm{X}\bm{P}-\bm{X}_{\star}\right\Vert _{\mathrm{F}}^{2}+\left\Vert \bm{Y}\bm{P}^{-\top}-\bm{X}_{\star}\right\Vert _{\mathrm{F}}^{2}.
\]
Then we have 
\begin{equation*}
 \left\langle \bm{X}-\bm{X}_{\star},\left(\bm{X}\bm{Y}^{\top}-\bm{M}_{\star}\right)\bm{Y}\right\rangle 
 \geq\left\Vert \bm{Y}(\bm{X}-\bm{X}_{\star})^{\top}\right\Vert _{\mathrm{F}}^{2}-\frac{1}{4}\left\Vert \bm{X}-\bm{X}_{\star}\right\Vert _{\mathrm{F}}^{4},
\end{equation*}
and similarly,
\begin{equation*}
  \left\langle \bm{Y}-\bm{Y}_{\star},\left(\bm{X}\bm{Y}^{\top}-\bm{M}_{\star}\right)^{\top}\bm{X}\right\rangle 
 \geq\left\Vert \bm{X}(\bm{Y}-\bm{Y}_{\star})^{\top}\right\Vert _{\mathrm{F}}^{2}-\frac{1}{4}\left\Vert \bm{Y}-\bm{Y}_{\star}\right\Vert _{\mathrm{F}}^{4}.
\end{equation*}
\end{lemma}

\begin{lemma}[Smoothness]\label{lemma:smoothness}Suppose that
$\left\Vert \bm{Y}-\bm{Y}_{\star}\right\Vert \le\sigma_{1}\left(\bm{Y}_{\star}\right)/4$,
then one has 
\begin{equation*}
 \left\Vert \left(\bm{X}\bm{Y}^{\top}-\bm{M}_{\star}\right)\bm{Y}\right\Vert _{\mathrm{F}} \leq
  \frac{3}{2}\sigma_{1}(\bm{Y}_{\star})    \Big(\left\Vert (\bm{X}-\bm{X}_{\star})\bm{Y}^{\top}\right\Vert _{\mathrm{F}}+\left\Vert \bm{X}(\bm{Y}-\bm{Y}_{\star})^{\top}\right\Vert _{\mathrm{F}} 
 +\left\Vert \bm{X}-\bm{X}_{\star}\right\Vert _{\mathrm{F}}\left\Vert \bm{Y}-\bm{Y}_{\star}\right\Vert _{\mathrm{F}} \Big).
\end{equation*}
Similarly, with the proviso that $\left\Vert \bm{X}-\bm{X}_{\star}\right\Vert \le\sigma_{1}\left(\bm{X}_{\star}\right)/4$,
one has 
\begin{equation*}
 \left\Vert \left(\bm{X}\bm{Y}^{\top}-\bm{M}_{\star}\right)^{\top}\bm{X}\right\Vert _{\mathrm{F}}  \leq 
\frac{3}{2}\sigma_{1}(\bm{X}_{\star})  \Big(\left\Vert (\bm{X}-\bm{X}_{\star})\bm{Y}^{\top}\right\Vert _{\mathrm{F}}+\left\Vert \bm{X}(\bm{Y}-\bm{Y}_{\star})^{\top}\right\Vert _{\mathrm{F}} 
 + \left\Vert \bm{X}-\bm{X}_{\star}\right\Vert _{\mathrm{F}}\left\Vert \bm{Y}-\bm{Y}_{\star}\right\Vert _{\mathrm{F}} \Big).
\end{equation*}
\end{lemma}

\subsection{Proof of Theorem~\ref{thm:matrix_factorization} \label{subsec:Proof-of-Theorem_MF}}

With the help of Lemmas~\ref{lemma:alignment-near-rotation}--\ref{lemma:smoothness},
we are in a position to establish Theorem~\ref{thm:matrix_factorization}.
Denote by $\hat{\bm{R}}\in\mathbb{R}^{r\times r}$ the best rotation
matrix between $\bm{Z}_{0}$ and $\bm{Z}_{\star}$, that is 
\[
\hat{\bm{R}}\triangleq\argmin_{\bm{R}\in\mathcal{O}^{r\times r}}\left\Vert \bm{Z}_{0}\bm{R}-\bm{Z}_{\star}\right\Vert _{\mathrm{F}}.
\]
Combine the assumption of initialization (cf.~(\ref{eq:init-MF}))
and Lemma~\ref{lemma:alignment-near-rotation} to see that 
\[
\bm{Q}_{0}\triangleq\argmin_{\substack{\bm{P}\in\mathbb{R}^{r\times r} \\ \text{invertible}}}  \left\Vert \bm{X}_{0}\bm{P}-\bm{X}_{\star}\right\Vert _{\mathrm{F}}^{2}+\left\Vert \bm{Y}_{0}\bm{P}^{-\top}-\bm{Y}_{\star}\right\Vert _{\mathrm{F}}^{2}
\]
exists and in addition, one has 
\[
\Vert\bm{Q}_{0}-\hat{\bm{R}}\Vert\leq\frac{5c_{0}}{\kappa^{3/2}}\leq\frac{1}{400\sqrt{\kappa}}
\]
as long as $c_{0}>0$ is sufficiently small.

The remaining proof is inductive in nature. In particular, we aim
at proving the following induction hypotheses. 
\begin{enumerate}
\item The optimal alignment matrix $\bm{Q}_{t}$ between $\bm{Z}_{t}$ and
$\bm{Z}_{\star}$ exists. 
\item The distance between $\bm{Z}_{t}$ and $\bm{Z}_{\star}$ obeys 
\[
\mathrm{dist}\left(\bm{Z}_{t},\bm{Z}_{\star}\right)\leq\left(1-\frac{\eta}{50\kappa}\right)^{t}\mathrm{dist}\left(\bm{Z}_{0},\bm{Z}_{\star}\right).
\]
\item The optimal alignment matrix $\bm{Q}_{t}$ is nearly a rotation matrix
in the sense that 
\[
\Vert\bm{Q}_{t}-\hat{\bm{R}}\Vert\leq\frac{1}{400\sqrt{\kappa}}.
\]
\end{enumerate}
It is straightforward to check that these three claims hold for $t=0$.
In what follows, we shall assume that the induction hypotheses hold
for all iterations up to the $t$th iteration and intend to establish
that they continue to hold for the $(t+1)$th iteration.

\paragraph{Verifying the first induction hypothesis} We begin with demonstrating the existence of $\bm{Q}_{t+1}$. In view
of the gradient update rule (\ref{eq:gd-update}), we have 
\begin{align*}
\bm{X}_{t+1}\bm{Q}_{t} & =\bm{X}_{t}\bm{Q}_{t}-\frac{\eta}{\sigma_{\mathrm{\max}}}\left(\bm{X}_{t}\bm{Y}_{t}^{\top}-\bm{M}_{\star}\right)\bm{Y}_{t}\bm{Q}_{t} =\tilde{\bm{X}}_{t}-\frac{\eta}{\sigma_{\mathrm{\max}}}\left(\tilde{\bm{X}}_{t}\tilde{\bm{Y}}_{t}^{\top}-\bm{M}_{\star}\right)\tilde{\bm{Y}}_{t}\left(\bm{Q}_{t}^{\top}\bm{Q}_{t}\right),\\
\bm{Y}_{t+1}\bm{Q}_{t}^{-\top} & =\bm{Y}_{t}\bm{Q}_{t}^{-\top}-\frac{\eta}{\sigma_{\mathrm{max}}}\left(\bm{X}_{t}\bm{Y}_{t}^{\top}-\bm{M}_{\star}\right)^{\top}\bm{X}_{t}\bm{Q}_{t}^{-\top}  =\tilde{\bm{Y}}_{t}-\frac{\eta}{\sigma_{\mathrm{\max}}}\left(\bm{X}_{t}\bm{Y}_{t}^{\top}-\bm{M}_{\star}\right)^{\top}\tilde{\bm{X}}_{t}\left(\bm{Q}_{t}^{\top}\bm{Q}_{t}\right)^{-1},
\end{align*}
where we denote 
\[
\tilde{\bm{X}}_{t}\triangleq\bm{X}_{t}\bm{Q}_{t}\qquad\text{and}\qquad\tilde{\bm{Y}}_{t}\triangleq\bm{Y}_{t}\bm{Q}_{t}^{-\top}.
\]
As a result, one has the following equality 
\[
\begin{aligned} & \left\Vert \bm{X}_{t+1}\bm{Q}_{t}-\bm{X}_{\star}\right\Vert _{\mathrm{F}}^{2}+\left\Vert \bm{Y}_{t+1}\bm{Q}_{t}^{-\top}-\bm{Y}_{\star}\right\Vert _{\mathrm{F}}^{2}\\
 & \quad=\underbrace{\left\Vert \tilde{\bm{X}}_{t}-\bm{X}_{\star}-\frac{\eta}{\sigma_{\mathrm{\max}}}\left(\tilde{\bm{X}}_{t}\tilde{\bm{Y}}_{t}^{\top}-\bm{M}_{\star}\right)\tilde{\bm{Y}}_{t}\bm{\Lambda}_{t}\right\Vert _{\mathrm{F}}^{2}}_{\eqqcolon\alpha_{1}} +\underbrace{\left\Vert \tilde{\bm{Y}}_{t}-\bm{Y}_{\star}-\frac{\eta}{\sigma_{\mathrm{\max}}}\left(\tilde{\bm{X}}_{t}\tilde{\bm{Y}}_{t}^{\top}-\bm{M}_{\star}\right)^{\top}\tilde{\bm{X}}_{t}\bm{\Lambda}_{t}^{-1}\right\Vert _{\mathrm{F}}^{2}}_{\eqqcolon\alpha_{2}},
\end{aligned}
\]
where we have denoted $\bm{\Lambda}_{t}\triangleq\bm{Q}_{t}^{\top}\bm{Q}_{t}$.
By virtue of the third induction hypothesis, namely $\|\bm{Q}_{t}-\hat{\bm{R}}\|\leq{1}/({400\sqrt{\kappa}})$,
it is easy to check that $\|\bm{\Lambda}_{t}-\bm{I}_{r}\|\leq{1}/({180\sqrt{\kappa}})\triangleq\zeta$. Let
\[
\tilde{\bm{E}}_{\bm{X}_t}\triangleq \tilde{\bm{X}}_{t}-\bm{X}_{\star}\qquad\text{and}\qquad\tilde{\bm{E}}_{\bm{Y}_t}\triangleq\tilde{\bm{Y}}_{t}-\bm{Y}_{\star}.
\]
Expand $\alpha_{1}$ to obtain 
\begin{align*}
\alpha_{1}=& \big\Vert \tilde{\bm{E}}_{\bm{X}_t} \big\Vert _{\mathrm{F}}^{2}  +\left(\frac{\eta}{\sigma_{\mathrm{max}}}\right)^{2}\underbrace{\left\Vert \left(\tilde{\bm{X}}_{t}\tilde{\bm{Y}}_{t}^{\top}-\bm{M}_{\star}\right)\tilde{\bm{Y}_{t}}\bm{\Lambda}_{t}\right\Vert _{\mathrm{F}}^{2}}_{\eqqcolon\beta_{1}}  -2\frac{\eta}{\sigma_{\mathrm{max}}}\underbrace{\left\langle \tilde{\bm{E}}_{\bm{X}_t} ,\left(\tilde{\bm{X}}_{t}\tilde{\bm{Y}}_{t}^{\top}-\bm{M}_{\star}\right)\tilde{\bm{Y}_{t}}\bm{\Lambda}_{t}\right\rangle }_{\eqqcolon\gamma_{1}}.
\end{align*}
Similarly, we can decompose $\alpha_{2}$ into 
\[
\begin{aligned}\alpha_{2}=&\big\Vert \tilde{\bm{E}}_{\bm{Y}_t} \big\Vert _{\mathrm{F}}^{2}  +\left(\frac{\eta}{\sigma_{\mathrm{\max}}}\right)^{2}\underbrace{\left\Vert \left(\tilde{\bm{X}}_{t}\tilde{\bm{Y}}_{t}^{\top}-\bm{M}_{\star}\right)^{\top}\tilde{\bm{X}}_{t}\bm{\Lambda}_{t}^{-1}\right\Vert _{\mathrm{F}}^{2}}_{\eqqcolon\beta_{2}}  -2\frac{\eta}{\sigma_{\mathrm{max}}}\underbrace{\left\langle\tilde{\bm{E}}_{\bm{Y}_t} ,\left(\tilde{\bm{X}}_{t}\tilde{\bm{Y}}_{t}^{\top}-\bm{M}_{\star}\right)^{\top}\tilde{\bm{X}}_{t}\bm{\Lambda}_{t}^{-1}\right\rangle }_{\eqqcolon\gamma_{2}}.\end{aligned}
\]
We intend to apply Lemma~\ref{lemma:gradient-dominance} to lower
bound the terms $\gamma_{1}$ and $\gamma_{2}$ and apply Lemma~\ref{lemma:smoothness}
to upper bound $\beta_{1}$ and $\beta_{2}$. First, since $(\tilde{\bm{X}}_{t},\tilde{\bm{Y}}_{t})$
is aligned with $(\bm{X}_{\star},\bm{Y}_{\star})$, we can invoke
Lemma~\ref{lemma:gradient-dominance} to see that 
\begin{align}
\gamma_{1} & \geq\left\langle\tilde{\bm{E}}_{\bm{X}_t} ,\left(\tilde{\bm{X}}_{t}\tilde{\bm{Y}}_{t}^{\top}-\bm{M}_{\star}\right)\tilde{\bm{Y}_{t}}\right\rangle -\left|\left\langle \tilde{\bm{E}}_{\bm{X}_t},\left(\tilde{\bm{X}}_{t}\tilde{\bm{Y}}_{t}^{\top}-\bm{M}_{\star}\right)\tilde{\bm{Y}_{t}}\left(\bm{\Lambda}_{t}-\bm{I}_{r}\right)\right\rangle \right|\nonumber \\
 & \geq \big\Vert \tilde{\bm{Y}}_{t}\tilde{\bm{E}}_{\bm{X}_t}^{\top}\big\Vert _{\mathrm{F}}^{2}-\frac{1}{4}\big\Vert \tilde{\bm{E}}_{\bm{X}_t}\big\Vert _{\mathrm{F}}^{4} -\|\bm{\Lambda}_{t}-\bm{I}_{r}\|\cdot\left\Vert \left(\tilde{\bm{X}}_{t}\tilde{\bm{Y}}_{t}^{\top}-\bm{M}_{\star}\right)\tilde{\bm{Y}_{t}}\right\Vert _{\mathrm{F}}\big\Vert \tilde{\bm{E}}_{\bm{X}_t}\big\Vert _{\mathrm{F}}\nonumber \\
 & \geq\big\Vert \tilde{\bm{Y}}_{t}\tilde{\bm{E}}_{\bm{X}_t}^{\top}\big\Vert _{\mathrm{F}}^{2}-\frac{1}{400}\sigma_{\min}\big\Vert \tilde{\bm{E}}_{\bm{X}_t}\big\Vert _{\mathrm{F}}^{2} -\zeta\left\Vert \left(\tilde{\bm{X}}_{t}\tilde{\bm{Y}}_{t}^{\top}-\bm{M}_{\star}\right)\tilde{\bm{Y}_{t}}\right\Vert _{\mathrm{F}}\big\Vert \tilde{\bm{E}}_{\bm{X}_t}\big\Vert _{\mathrm{F}}.\label{eq:lower_bound_gamma_1}
\end{align}
Here the last line follows from the bound $\|\bm{\Lambda}_{t}-\bm{I}_{r}\|\leq\zeta$
and the second induction hypothesis, i.e. 
\[
\big\Vert \tilde{\bm{E}}_{\bm{X}_t}\big\Vert _{\mathrm{F}}^{2}\leq\mathrm{dist}^{2}\left(\bm{Z}_{t},\bm{Z}_{\star}\right)\leq\mathrm{dist}^{2}\left(\bm{Z}_{0},\bm{Z}_{\star}\right)\leq\frac{1}{100}\sigma_{\min}.
\]
The last term in (\ref{eq:lower_bound_gamma_1}) can be further bounded
via Lemma \ref{lemma:smoothness} as 
\begin{align*}
  \zeta\left\Vert \left(\tilde{\bm{X}}_{t}\tilde{\bm{Y}}_{t}^{\top}-\bm{M}_{\star}\right)\tilde{\bm{Y}}_{t}\right\Vert _{\mathrm{F}}\big\Vert\tilde{\bm{E}}_{\bm{X}_t}\big\Vert _{\mathrm{F}}& \leq\frac{3\zeta}{2}\sqrt{\sigma_{\max}}   \Big(\big\Vert\tilde{\bm{E}}_{\bm{X}_t}\tilde{\bm{Y}}_{t}^{\top}\big\Vert _{\mathrm{F}}+\big\Vert \tilde{\bm{X}}_{t}\tilde{\bm{E}}_{\bm{Y}_t}^{\top}\big\Vert _{\mathrm{F}} +\big\Vert\tilde{\bm{E}}_{\bm{X}_t} \big\Vert _{\mathrm{F}}\big\Vert \tilde{\bm{E}}_{\bm{Y}_t}\big\Vert _{\mathrm{F}}\Big)\big\Vert \tilde{\bm{E}}_{\bm{X}_t} \big\Vert _{\mathrm{F}}\\
 &  =\frac{9\zeta\sqrt{\kappa}}{2}\big\Vert \tilde{\bm{E}}_{\bm{X}_t}\tilde{\bm{Y}}_{t}^{\top}\big\Vert _{\mathrm{F}}\cdot\frac{\sqrt{\sigma_{\min}}}{3}\big\Vert\tilde{\bm{E}}_{\bm{X}_t}\big\Vert _{\mathrm{F}} +\frac{9\zeta\sqrt{\kappa}}{2}\big\Vert \tilde{\bm{X}}_{t}\tilde{\bm{E}}_{\bm{Y}_t}^{\top}\big\Vert _{\mathrm{F}}\cdot\frac{\sqrt{\sigma_{\min}}}{3}\big\Vert\tilde{\bm{E}}_{\bm{X}_t}\big\Vert _{\mathrm{F}}\\
 & \qquad\qquad+\frac{3\zeta}{2}\sqrt{\sigma_{\max}}\big\Vert \tilde{\bm{E}}_{\bm{Y}_t}\big\Vert _{\mathrm{F}}\big\Vert\tilde{\bm{E}}_{\bm{X}_t}\big\Vert _{\mathrm{F}}^{2}\\
 &  \leq\frac{81\zeta^{2}\kappa}{8}\big\Vert \tilde{\bm{E}}_{\bm{X}_t}\tilde{\bm{Y}}_{t}^{\top}\big\Vert _{\mathrm{F}}^{2}+\frac{81\zeta^{2}\kappa}{8}\big\Vert \tilde{\bm{X}}_{t} \tilde{\bm{E}}_{\bm{Y}_t}^{\top}\big\Vert _{\mathrm{F}}^{2} +\frac{\sigma_{\min}}{8}\big\Vert\tilde{\bm{E}}_{\bm{X}_t}\big\Vert _{\mathrm{F}}^{2},
\end{align*}
where the last inequality arises since $ab\leq(a^{2}+b^{2})/2$ and
\begin{align*}
\frac{3\zeta}{2}\sqrt{\sigma_{\max}}\big\Vert \tilde{\bm{E}}_{\bm{Y}_t}\big\Vert _{\mathrm{F}} & \leq\frac{3\zeta}{2}\sqrt{\sigma_{\max}}\,\mathrm{dist}\left(\bm{Z}_{0},\bm{Z}_{\star}\right) \\
& \leq\frac{3\zeta}{2}\sqrt{\sigma_{\max}}c_{0}\frac{1}{\kappa^{3/2}}\sqrt{\sigma_{\min}}\leq\frac{\sigma_{\min}}{72}
\end{align*}
as long as $c_{0}$ is sufficiently small. Combine the above two bounds
to reach 
\begin{align*}
\gamma_{1}& \geq\left(1-\frac{81\zeta^{2}\kappa}{8}\right) \big\Vert \tilde{\bm{Y}}_{t}\tilde{\bm{E}}_{\bm{X}_t}^{\top} \big\Vert _{\mathrm{F}}^{2}   -\frac{81\zeta^{2}\kappa}{8} \big\Vert \tilde{\bm{X}}_{t}\tilde{\bm{E}}_{\bm{Y}_t}^{\top} \big\Vert _{\mathrm{F}}^{2}-\frac{\sigma_{\min}}{7}\big\Vert\tilde{\bm{E}}_{\bm{X}_t}\big\Vert _{\mathrm{F}}^{2}.
\end{align*}
Similarly, $\gamma_{2}$ can be lower bounded as 
\begin{align*}
\gamma_{2}& \geq\left(1-\frac{81\zeta^{2}\kappa}{8}\right)\big\Vert \tilde{\bm{X}}_{t}\tilde{\bm{E}}_{\bm{Y}_t}^{\top}\big\Vert _{\mathrm{F}}^{2}   -\frac{81\zeta^{2}\kappa}{8} \big\Vert \tilde{\bm{Y}}_{t}\tilde{\bm{E}}_{\bm{X}_t}^{\top}\big\Vert _{\mathrm{F}}^{2}-\frac{\sigma_{\min}}{7}\big\Vert \tilde{\bm{E}}_{\bm{Y}_t} \big\Vert _{\mathrm{F}}^{2},
\end{align*}
which together with the bound on $\gamma_{1}$ implies 
\begin{align*}
\gamma_{1}+\gamma_{2} 
& \geq\left(1-\frac{81\zeta^{2}\kappa}{4}\right)\left(\big\Vert \tilde{\bm{Y}}_{t}\tilde{\bm{E}}_{\bm{X}_t}^{\top}\big\Vert _{\mathrm{F}}^{2}+\big\Vert \tilde{\bm{X}}_{t} \tilde{\bm{E}}_{\bm{Y}_t}^{\top}\big\Vert _{\mathrm{F}}^{2}\right) -\frac{\sigma_{\min}}{7}\mathrm{dist}^{2}\left(\bm{Z}_{t},\bm{Z}_{\star}\right)\\
 & \geq\frac{3}{4}\left(\big\Vert \tilde{\bm{Y}}_{t}\tilde{\bm{E}}_{\bm{X}_t}^{\top}\big\Vert _{\mathrm{F}}^{2}+\big\Vert \tilde{\bm{X}}_{t}\tilde{\bm{E}}_{\bm{Y}_t}^{\top}\big\Vert _{\mathrm{F}}^{2}\right)-\frac{\sigma_{\min}}{7}\mathrm{dist}^{2}\left(\bm{Z}_{t},\bm{Z}_{\star}\right),
\end{align*}
where we plug in the definition of $\zeta=1/(180\sqrt{\kappa})$.

Now we move on to controlling $\beta_{1}$ and $\beta_{2}$. Recognizing
that $\|\bm{\Lambda}_{t}\|\leq2$, one has 
\begin{align}
\beta_{1} & \leq4\left\Vert \left(\tilde{\bm{X}}_{t}\tilde{\bm{Y}}_{t}^{\top}-\bm{M}_{\star}\right)\tilde{\bm{Y}}_{t}\right\Vert _{\mathrm{F}}^{2}\nonumber \\
 & \leq9\sigma_{\max}\Big(\big\Vert \tilde{\bm{E}}_{\bm{X}_t}\tilde{\bm{Y}}_{t}^{\top}\big\Vert _{\mathrm{F}}+\big\Vert \tilde{\bm{X}}_{t}\tilde{\bm{E}}_{\bm{Y}_t}^{\top}\big\Vert _{\mathrm{F}} +\big\Vert\tilde{\bm{E}}_{\bm{X}_t}\big\Vert _{\mathrm{F}}\big\Vert \tilde{\bm{E}}_{\bm{Y}_t} \big\Vert _{\mathrm{F}}\Big)^{2},\label{eq:upper_beta_1}
\end{align}
where the second line follows from Lemma~\ref{lemma:smoothness}.
Apply the elementary inequality $(a+b+c)^{2}\leq3(a^{2}+b^{2}+c^{2})$
to see that 
\begin{align*}
\beta_{1} & \leq27\sigma_{\max}\Big(  \big\Vert \tilde{\bm{E}}_{\bm{X}_t}\tilde{\bm{Y}}_{t}^{\top}\big\Vert _{\mathrm{F}}^{2}+\big\Vert \tilde{\bm{X}}_{t}\tilde{\bm{E}}_{\bm{Y}_t}^{\top}\big\Vert _{\mathrm{F}}^{2}  +\big\Vert\tilde{\bm{E}}_{\bm{X}_t}\big\Vert _{\mathrm{F}}^{2}\big\Vert \tilde{\bm{E}}_{\bm{Y}_t}\big\Vert _{\mathrm{F}}^{2}\Big) \\
 & \leq 27\sigma_{\max} \Big(\big\Vert \tilde{\bm{E}}_{\bm{X}_t}\tilde{\bm{Y}}_{t}^{\top}\big\Vert _{\mathrm{F}}^{2}+\big\Vert \tilde{\bm{X}}_{t}\tilde{\bm{E}}_{\bm{Y}_t}^{\top}\big\Vert _{\mathrm{F}}^{2} \Big) +27c_{0}^{2}\frac{\sigma_{\max}\sigma_{\min}}{\kappa^{3}}\big\Vert\tilde{\bm{E}}_{\bm{X}_t}\big\Vert _{\mathrm{F}}^{2}.
\end{align*}
Here the second line relies on the fact that $\|\tilde{\bm{E}}_{\bm{Y}_t}\|_{\mathrm{F}}^{2}\leq\text{dist}^{2}(\bm{Z}_{0},\bm{Z}_{\star})\leq c_{0}^{2}\sigma_{\min}/\kappa^{3}$.
Similarly, one can bound $\beta_{2}$ as 
\begin{align*}
\beta_{2}& \leq27\sigma_{\max}\left(\big\Vert \tilde{\bm{E}}_{\bm{X}_t}\tilde{\bm{Y}}_{t}^{\top}\big\Vert _{\mathrm{F}}^{2}+\big\Vert \tilde{\bm{X}}_{t}\tilde{\bm{E}}_{\bm{Y}_t}^{\top}\big\Vert _{\mathrm{F}}^{2}\right)  +27c_{0}^{2}\frac{\sigma_{\max}\sigma_{\min}}{\kappa^{3}}\big\Vert \tilde{\bm{E}}_{\bm{Y}_t} \big\Vert _{\mathrm{F}}^{2},
\end{align*}
which in conjunction with the bound on $\beta_{1}$ yields 
\begin{align*}
\beta_{1}+\beta_{2}\leq & 54\sigma_{\max}\left(\big\Vert \tilde{\bm{E}}_{\bm{X}_t}\tilde{\bm{Y}}_{t}^{\top}\big\Vert _{\mathrm{F}}^{2}+\big\Vert \tilde{\bm{X}}_{t}\tilde{\bm{E}}_{\bm{Y}_t}^{\top}\big\Vert _{\mathrm{F}}^{2}\right)  +27c_{0}^{2}\frac{\sigma_{\max}\sigma_{\min}}{\kappa^{3}}\mathrm{dist}^{2}\left(\bm{Z}_{t},\bm{Z}_{\star}\right).
\end{align*}

Collect all the bounds on $\alpha_{1}$ and $\alpha_{2}$ to arrive
at 
\begin{align*}
 & \left\Vert \bm{X}_{t+1}\bm{Q}_{t}-\bm{X}_{\star}\right\Vert _{\mathrm{F}}^{2}+\left\Vert \bm{Y}_{t+1}\bm{Q}_{t}^{-\top}-\bm{Y}_{\star}\right\Vert _{\mathrm{F}}^{2}\\
 & \quad\leq\left(1+\frac{27c_{0}^{2}\eta^{2}}{\kappa^{4}}\right)\mathrm{dist}^{2}\left(\bm{Z}_{t},\bm{Z}_{\star}\right) +\left(\frac{54\eta^{2}}{\sigma_{\mathrm{max}}}\right)\left(\big\Vert \tilde{\bm{E}}_{\bm{X}_t}\tilde{\bm{Y}}_{t}^{\top}\big\Vert _{\mathrm{F}}^{2}+\big\Vert \tilde{\bm{X}}_{t}\tilde{\bm{E}}_{\bm{Y}_t}^{\top}\big\Vert _{\mathrm{F}}^{2}\right)\\
 & \qquad\qquad-2\frac{\eta}{\sigma_{\max}}\left[\frac{3}{4}\Big(\big\Vert \tilde{\bm{E}}_{\bm{X}_t} \tilde{\bm{Y}}_{t}^{\top}\big\Vert _{\mathrm{F}}^{2}+\big\Vert \tilde{\bm{X}}_{t}\tilde{\bm{E}}_{\bm{Y}_t}^{\top}\big\Vert _{\mathrm{F}}^{2}\Big) -\frac{\sigma_{\min}}{7}\mathrm{dist}^{2}\left(\bm{Z}_{t},\bm{Z}_{\star}\right)\right]\\
 & \quad=\left(1+\frac{27c_{0}^{2}\eta^{2}}{\kappa^{4}}+\frac{\eta}{3.5\kappa}\right)\mathrm{dist}^{2}\left(\bm{Z}_{t},\bm{Z}_{\star}\right) +\left(\frac{54\eta^{2}}{\sigma_{\max}}-\frac{3\eta}{2\sigma_{\max}}\right)\cdot\left(\big\Vert \tilde{\bm{E}}_{\bm{X}_t}\tilde{\bm{Y}}_{t}^{\top}\big\Vert _{\mathrm{F}}^{2}+\big\Vert \tilde{\bm{X}}_{t}\tilde{\bm{E}}_{\bm{Y}_t}^{\top}\big\Vert _{\mathrm{F}}^{2}\right)\\
 & \quad\leq\left(1+\frac{\eta}{3\kappa}\right)\mathrm{dist}^{2}\left(\bm{Z}_{t},\bm{Z}_{\star}\right)  -\frac{3\eta}{4\sigma_{\max}} \left(\sigma_{r}^{2}(\tilde{\bm{Y}}_{t})\big\Vert\tilde{\bm{E}}_{\bm{X}_t}\big\Vert _{\mathrm{F}}^{2}+\sigma_{r}^{2}(\tilde{\bm{X}}_{t})\big\Vert \tilde{\bm{E}}_{\bm{Y}_t}\big\Vert _{\mathrm{F}}^{2}\right),
\end{align*}
where the last line follows as long as $\eta\leq1/24$. Furthermore,
since $\sigma_{r}^{2}(\tilde{\bm{Y}}_{t})\geq\sigma_{\min}/2$ and
$\sigma_{r}^{2}(\tilde{\bm{X}}_{t})\geq\sigma_{\min}/2$, we have
\begin{align}
\left\Vert \bm{X}_{t+1}\bm{Q}_{t}-\bm{X}_{\star}\right\Vert _{\mathrm{F}}^{2}  +\left\Vert \bm{Y}_{t+1}\bm{Q}_{t}^{-\top}-\bm{Y}_{\star}\right\Vert _{\mathrm{F}}^{2}  &\leq\left(1-\frac{\eta}{24\kappa}\right)\mathrm{dist}^{2}\left(\bm{Z}_{t},\bm{Z}_{\star}\right).\label{eq:distance-inequality}
\end{align}
Lemma~\ref{lemma:alignment-near-rotation} then ensures the existence
of $\bm{Q}_{t+1}$. 

\paragraph{Verifying the second induction hypothesis} The second induction hypothesis for the $(t+1)$th iteration follows
immediately from the above proof. Since $\bm{Q}_{t+1}$ exists, by
definition, one has 
\begin{align*}
 \mathrm{dist}\left(\bm{Z}_{t+1},\bm{Z}_{\star}\right) 
&  =\sqrt{\left\Vert \bm{X}_{t+1}\bm{Q}_{t+1}-\bm{X}_{\star}\right\Vert _{\mathrm{F}}^{2}+\left\Vert \bm{Y}_{t+1}\bm{Q}_{t+1}^{-\top}-\bm{X}_{\star}\right\Vert _{\mathrm{F}}^{2}}\\
 &  \leq\sqrt{\left\Vert \bm{X}_{t+1}\bm{Q}_{t}-\bm{X}_{\star}\right\Vert _{\mathrm{F}}^{2}+\left\Vert \bm{Y}_{t+1}\bm{Q}_{t}^{-\top}-\bm{X}_{\star}\right\Vert _{\mathrm{F}}^{2}}\\
 &  \leq\left(1-\frac{\eta}{50\kappa}\right)\mathrm{dist}\left(\bm{Z}_{t},\bm{Z}_{\star}\right).
\end{align*}

\paragraph{Verifying the third induction hypothesis} It remains to show the last induction hypothesis, namely $\|\bm{Q}_{t+1}-\hat{\bm{R}}\|\leq{1}/({400\sqrt{\kappa}})$. In view of (\ref{eq:distance-inequality}), one has $\max\{\Vert\bm{X}_{t+1}\bm{Q}_{t}-\bm{X}_{\star}\Vert_{\mathrm{F}},\Vert\bm{Y}_{t+1}\bm{Q}_{t}^{-\top}-\bm{Y}_{\star}\Vert_{\mathrm{F}}\}\leq\mathrm{dist}\left(\bm{Z}_{t},\bm{Z}_{\star}\right)$.
Invoke Lemma~\ref{lemma:alignment-near-rotation} again to arrive
at 
\begin{align*}
\left\Vert \bm{Q}_{t+1}-\bm{Q}_{t}\right\Vert  & \leq\frac{5}{\sigma_{r}\left(\bm{X}_{\star}\right)}\mathrm{dist}\left(\bm{Z}_{t},\bm{Z}_{\star}\right)\\
 & \leq\frac{5}{\sigma_{r}\left(\bm{X}_{\star}\right)}\left(1-\frac{\eta}{50\kappa}\right)^{t}c_{0}\frac{1}{\kappa^{3/2}}\sigma_{r}\left(\bm{X}_{\star}\right)\\
 & \leq5c_{0}\left(1-\frac{\eta}{50\kappa}\right)^{t}\frac{1}{\kappa^{3/2}}.
\end{align*}
Hence, by the triangle inequality and the telescoping sum, we obtain
\begin{align*}
\big\Vert \bm{Q}_{t+1}-\hat{\bm{R}}\big\Vert  & \leq\sum_{s=0}^{t}\left\Vert \bm{Q}_{s+1}-\bm{Q}_{s}\right\Vert +\big\Vert \bm{Q}_{0}-\hat{\bm{R}}\big\Vert \\
 & \leq5c_{0}\sum_{s=0}^{t}\left(1-\frac{\eta}{50\kappa}\right)^{s}\frac{1}{\kappa^{3/2}}+5c_{0}\frac{1}{\kappa^{3/2}}\\
 & <5c_{0}\sum_{s=0}^{\infty}\left(1-\frac{\eta}{50\kappa}\right)^{s}\frac{1}{\kappa^{3/2}}+5c_{0}\frac{1}{\kappa^{3/2}}\\
 & =5c_{0}\frac{50\kappa}{\eta}\frac{1}{\kappa^{3/2}}+5c_{0}\frac{1}{\kappa^{3/2}}\\
 & \leq\frac{1}{400\sqrt{\kappa}},
\end{align*}
as long as $c_{0}$ is small enough and $\eta$ is some constant.

Putting everything together, we finish the induction step and the proof is then completed. 


\subsection{Analysis for matrix sensing}

We now extend the techniques used in the proof of Theorem~\ref{thm:matrix_factorization}
to the matrix sensing case by leveraging the RIP. Suppose that the
initialization $\bm{Z}_{0}$ satisfies the condition~\eqref{eq:init_basin}.
By a standard argument as in \cite{tu2015low,zheng2016convergence,li2017nonconvex},\footnote{Since (i) the initialization $\bm{Z}_{0}$ is close to the ground truth $\bm{Z}_{\star}$, (ii) $\bm{X}_{0}$ and $\bm{Y}_{0}$ are balanced, it is obvious that the operator norm $\|\bm{X}_{0}\|^2 = \|\bm{Y}_{0}\|^2$ is orderwise equivalent to $\sigma_{\max}$. Therefore, all the convergence claims on using $\sigma_{\max}$ can be translated to those on using $\|\bm{X}_{0}\|^2$ and $\|\bm{Y}_{0}\|^2$ by adjusting $\eta$ up to some absolute constant. } it is sufficient to
consider the following update rule: 
\begin{equation}
\begin{aligned}\bX_{t+1} & =\bX_{t}-\frac{\eta}{\sigma_{\max}}\left[\mathcal{A}^{*}\mathcal{A}(\bX_{t}\bY_{t}^{\top}-\bm{M}_{\star})\right]\bY_{t};\\
\bY_{t+1} & =\bY_{t}-\frac{\eta}{\sigma_{\max}}\left[\mathcal{A}^{*}\mathcal{A}(\bX_{t}\bY_{t}^{\top}-\bm{M}_{\star})\right]^{\top}\bX_{t}.
\end{aligned}
\label{eq:sensing-gd-update}
\end{equation}
Compared with the update rule~\eqref{eq:gd-update} for low-rank
matrix factorization, the update rule for matrix sensing differs by
the operation of $\mathcal{A}^{*}\mathcal{A}$ when forming the gradient.
Therefore, we expect that GD has similar behaviors as earlier as long
as the operator $\mathcal{A}^{*}\mathcal{A}$ behaves as a near isometry
on low-rank matrices. This can be supplied by the following consequence
of the RIP. \begin{lemma}\label{lemma_rip_distance} Suppose that
$\mathcal{A}$ satisfies $2r$-RIP with a constant $\delta_{2r}$.
Then, for all matrices $\bM_{1}$ and $\bM_{2}$ of rank at most $r$,
we have 
\[
\left\vert \langle\mathcal{A}(\bM_{1}),\mathcal{A}(\bM_{2})\rangle-\langle\bM_{1},\bM_{2}\rangle\right\vert \le\delta_{2r}\left\Vert \bM_{1}\right\Vert _{\F}\left\Vert \bM_{2}\right\Vert _{\F}.
\]
Equivalently, we can write this as 
\[
\left|\mathrm{Tr}\left[\left(\mathcal{A}^{*}\mathcal{A}-\mathcal{I}\right)\left(\bm{M}_{1}\right)\bm{M}_{2}^{\top}\right]\right|\leq\delta_{2r}\left\Vert \bM_{1}\right\Vert _{\F}\left\Vert \bM_{2}\right\Vert _{\F}.
\]
A simple consequence is that for any $\bm{A}\in\mathbb{R}^{n_{2}\times r}$
\[
\left\Vert \left(\mathcal{A}^{*}\mathcal{A}-\mathcal{I}\right)\left(\bm{M}_{1}\right)\bm{A}\right\Vert _{\mathrm{F}}\leq\delta_{2r}\left\Vert \bM_{1}\right\Vert _{\F}\left\Vert \bm{A}\right\Vert .
\]
\end{lemma}

Similar to before, we denote $\tilde{\bm{X}}_{t}=\bm{X}_{t}\bm{Q}_{t}$
and $\tilde{\bm{Y}}_{t}=\bm{Y}_{t}\bm{Q}_{t}^{-\top}$, which are
aligned with $(\bm{X}_{\star},\bm{Y}_{\star})$. With this notation
in place, we can rewrite the update rule as 
\[
\begin{aligned}\bX_{t+1}\bQ_{t} & =\tilde{\bX}_{t}-\frac{\eta}{\sigma_{\max}}\left[\mathcal{A}^{*}\mathcal{A}(\bX_{t}\bY_{t}^{\top}-\bm{M}_{\star})\right]\tilde{\bY}_{t}\bm{\Lambda}_{t},\\
\bY_{t+1}\bQ_{t}^{-\top} & =\tilde{\bY}_{t}-\frac{\eta}{\sigma_{\max}}\left[\mathcal{A}^{*}\mathcal{A}(\bX_{t}\bY_{t}^{\top}-\bm{M}_{\star})\right]^{\top}\tilde{\bX}_{t}\bm{\Lambda}_{t}^{-1}.
\end{aligned}
\]
where we recall $\bm{\Lambda}_{t}=\bm{Q}_{t}^{\top}\bm{Q}_{t}$. By
the definition of the distance function, we further obtain 
\begin{align}
 &\quad \mathrm{dist}^{2}(\bZ_{t+1},\bZ_{\star}) \nonumber\\
 &\le\left\Vert \bX_{t+1}\bQ_{t}-\bX_{\star}\right\Vert _{\F}^{2}+\left\Vert \bY_{t+1}\bQ_{t}^{-\top}-\bY_{\star}\right\Vert _{\F}^{2}\nonumber \\
 & =\left\Vert \tilde{\bX}_{t}-\frac{\eta}{\sigma_{\max}}\left[\mathcal{A}^{*}\mathcal{A}(\bX_{t}\bY_{t}^{\top}-\bm{M}_{\star})\right]\tilde{\bY}_{t}\bm{\Lambda}_{t}-\bX_{\star}\right\Vert _{\F}^{2}  +\left\Vert \tilde{\bY}_{t}-\frac{\eta}{\sigma_{\max}}\left[\mathcal{A}^{*}\mathcal{A}(\bX_{t}\bY_{t}^{\top}-\bm{M}_{\star})\right]^{\top}\tilde{\bX}_{t}\bm{\Lambda}_{t}^{-1}-\bY_{\star}\right\Vert _{\F}^{2}\nonumber \\
 & =\big\Vert \tilde{\bm{E}}_{\bm{X}_t}\big\Vert _{\F}^{2}+\big\Vert \tilde{\bm{E}}_{\bm{Y}_t}\big\Vert _{\F}^{2}\nonumber \\
 & \quad\quad+\left(\frac{\eta}{\sigma_{\max}}\right)^{2}\Big(\underbrace{\left\Vert \left[\mathcal{A}^{*}\mathcal{A}(\bX_{t}\bY_{t}^{\top}-\bm{M}_{\star})\right]\tilde{\bY}_{t}\bm{\Lambda}_{t}\right\Vert _{\F}^{2}}_{\eqqcolon\tilde{\beta}_{1}}  +\underbrace{\left\Vert \left[\mathcal{A}^{*}\mathcal{A}(\bX_{t}\bY_{t}^{\top}-\bm{M}_{\star})\right]^{\top}\tilde{\bX}_{t}\bm{\Lambda}_{t}^{-1}\right\Vert _{\F}^{2}}_{\eqqcolon\tilde{\beta}_{2}}\Big)\nonumber \\
 & \quad\quad-\frac{2\eta}{\sigma_{\max}}\Big(\underbrace{\left\langle \tilde{\bm{E}}_{\bm{X}_t},\left[\mathcal{A}^{*}\mathcal{A}(\bX_{t}\bY_{t}^{\top}-\bm{M}_{\star})\right]\tilde{\bY}_{t}\bm{\Lambda}_{t}\right\rangle }_{\eqqcolon\tilde{\gamma}_{1}} +\underbrace{\left\langle \tilde{\bm{E}}_{\bm{Y}_t},\left[\mathcal{A}^{*}\mathcal{A}(\bX_{t}\bY_{t}^{\top}-\bm{M}_{\star})\right]^{\top}\tilde{\bX}_{t}\bm{\Lambda}_{t}^{-1}\right\rangle }_{\eqqcolon\tilde{\gamma}_{2}}\Big),\label{equ_gd_induction}
\end{align}
where $\tilde{\bm{E}}_{\bm{X}_t}\triangleq \tilde{\bm{X}}_{t}-\bm{X}_{\star}$ and $\tilde{\bm{E}}_{\bm{Y}_t}\triangleq\tilde{\bm{Y}}_{t}-\bm{Y}_{\star}$. From the high level, the four terms $\tilde{\beta}_{1},\tilde{\beta}_{2},\tilde{\gamma}_{1}$
and $\tilde{\gamma}_{2}$ are the perturbed versions of $\beta_{1},\beta_{2},\gamma_{1}$
and $\gamma_{2}$ in Section~\ref{subsec:Proof-of-Theorem_MF}, respectively. 

For the first term, we have 
\begin{align}
  \sqrt{\tilde{\beta}_{1}}-\sqrt{\beta_{1}}   
& \overset{(\text{i})}{\leq}\left\Vert \left[\mathcal{A}^{*}\mathcal{A}\left(\bX_{t}\bY_{t}^{\top}-\bm{M}_{\star}\right)\right]\tilde{\bY}_{t}\bm{\Lambda}_{t}-\left(\bX_{t}\bY_{t}^{\top}-\bm{M}_{\star}\right)\tilde{\bY}_{t}\bm{\Lambda}_{t}\right\Vert _{\mathrm{F}}\nonumber \\
 & =\left\Vert \left[\left(\mathcal{A}^{*}\mathcal{A}-\mathcal{I}\right)\left(\bX_{t}\bY_{t}^{\top}-\bm{M}_{\star}\right)\right]\tilde{\bY}_{t}\bm{\Lambda}_{t}\right\Vert _{\mathrm{F}}\nonumber \\
 & \overset{(\text{ii})}{\leq}\left\Vert \left[\left(\mathcal{A}^{*}\mathcal{A}-\mathcal{I}\right) \tilde{\bm{E}}_{\bm{X}_t} \tilde{\bm{Y}}_{t}^{\top}\right]\tilde{\bY}_{t}\bm{\Lambda}_{t}\right\Vert _{\mathrm{F}} +\left\Vert \left[\left(\mathcal{A}^{*}\mathcal{A}-\bm{I}\right)\tilde{\bm{X}}_{t}\tilde{\bm{E}}_{\bm{Y}_t}^{\top}\right]\tilde{\bY}_{t}\bm{\Lambda}_{t}\right\Vert _{\mathrm{F}} +\left\Vert \left[\left(\mathcal{A}^{*}\mathcal{A}-\bm{I}\right) \tilde{\bm{E}}_{\bm{X}_t} \tilde{\bm{E}}_{\bm{Y}_t}^{\top}\right]\tilde{\bY}_{t}\bm{\Lambda}_{t}\right\Vert _{\mathrm{F}}\nonumber \\
 & \overset{(\text{iii})}{\leq}\delta_{2r}\left( \big\Vert \tilde{\bm{E}}_{\bm{X}_t}\tilde{\bm{Y}}_{t}^{\top}\big\Vert_{\mathrm{F}}+\big\Vert \tilde{\bm{X}}_{t} \tilde{\bm{E}}_{\bm{Y}_t}^{\top}\big\Vert_{\mathrm{F}}+\big\Vert \tilde{\bm{E}}_{\bm{X}_t} \tilde{\bm{E}}_{\bm{Y}_t}^{\top}\big\Vert_{\mathrm{F}}\right) \big\Vert \tilde{\bY}_{t}\bm{\Lambda}_{t}\big\Vert \nonumber \\
 & \overset{(\text{iv})}{\leq}4\delta_{2r}\sqrt{\sigma_{\max}}\left(\big\Vert  \tilde{\bm{E}}_{\bm{X}_t} \tilde{\bm{Y}}_{t}^{\top}\big\Vert_{\text{F}}+\big\Vert \tilde{\bm{X}}_{t} \tilde{\bm{E}}_{\bm{Y}_t}^{\top}\big\Vert_{\mathrm{F}}+\big\Vert  \tilde{\bm{E}}_{\bm{X}_t} \tilde{\bm{E}}_{\bm{Y}_t}^{\top}\big\Vert_{\mathrm{F}}\right).\label{eq:diff_beta_tilde}
\end{align}
Here, the first (i) and second (ii) inequalities follow from the triangle
inequality. The third one (iii) uses Lemma~\ref{lemma_rip_distance}
and the last relation (iv) depends on $\|\bm{\Lambda}_{t}\|\leq2$
and $\|\tilde{\bm{Y}}_{t}\|\leq2\sqrt{\sigma_{\max}}$. Comparing
(\ref{eq:diff_beta_tilde}) with (\ref{eq:upper_beta_1}) reveals
that $\tilde{\beta}_{1}-\beta_{1}$ constitutes a small perturbation
to $\beta_{1}$ when $\delta_{2r}$ is small. Similar bounds hold
for $\sqrt{\tilde{\beta}_{2}}-\sqrt{\beta_{2}}$. As a result, when $\delta_{2r}$
is sufficiently small, we have 
\begin{equation*}
\tilde{\beta}_{1}+\tilde{\beta}_{2}  \leq108\sigma_{\max}\left(\big\Vert \tilde{\bm{E}}_{\bm{X}_t} \tilde{\bm{Y}}_{t}^{\top}\big\Vert_{\mathrm{F}}^{2}+\big\Vert \tilde{\bm{X}}_{t}\tilde{\bm{E}}_{\bm{Y}_t}^{\top}\big\Vert_{\mathrm{F}}^{2}\right)
 +54c_{0}^{2}\frac{\sigma_{\max}\sigma_{\min}}{\kappa^{3}}\mathrm{dist}^{2}\left(\bm{Z}_{t},\bm{Z}_{\star}\right).
\end{equation*}

We now proceed to $\tilde{\gamma}_{1}$, for which we have 
\begin{align*}
 \left|\tilde{\gamma}_{1}-\gamma_{1}\right| 
&  =\Big|\left\langle \tilde{\bm{E}}_{\bm{X}_t},\left[\mathcal{A}^{*}\mathcal{A}\left(\bX_{t}\bY_{t}^{\top}-\bm{M}_{\star}\right)\right]\tilde{\bY}_{t}\bm{\Lambda}_{t}\right\rangle -\left\langle \tilde{\bm{E}}_{\bm{X}_t},\left(\bX_{t}\bY_{t}^{\top}-\bm{M}_{\star}\right)\tilde{\bY}_{t}\bm{\Lambda}_{t}\right\rangle \Big|\\
 &  =\left|\left\langle \tilde{\bm{E}}_{\bm{X}_t},\left[\left(\mathcal{A}^{*}\mathcal{A}-\mathcal{I}\right)\left(\bX_{t}\bY_{t}^{\top}-\bm{M}_{\star}\right)\right]\tilde{\bY}_{t}\bm{\Lambda}_{t}\right\rangle \right|\\
 &  =\left|\mathrm{Tr}\left(\left[\left(\mathcal{A}^{*}\mathcal{A}-\mathcal{I}\right)\left(\bX_{t}\bY_{t}^{\top}-\bm{M}_{\star}\right)\right]\tilde{\bY}_{t}\bm{\Lambda}_{t}\tilde{\bm{E}}_{\bm{X}_t}^{\top}\right)\right|\\
 &   \leq\left|\mathrm{Tr}\left(\left[\left(\mathcal{A}^{*}\mathcal{A}-\mathcal{I}\right)\left( \tilde{\bm{E}}_{\bm{X}_t}\tilde{\bm{Y}}_{t}^{\top}\right)\right]\tilde{\bY}_{t}\bm{\Lambda}_{t} \tilde{\bm{E}}_{\bm{X}_t}^{\top}\right)\right|+\left|\mathrm{Tr}\left(\left[\left(\mathcal{A}^{*}\mathcal{A}-\mathcal{I}\right)\left(\tilde{\bm{X}}_{t}\tilde{\bm{E}}_{\bm{Y}_t}^{\top}\right)\right]\tilde{\bY}_{t}\bm{\Lambda}_{t} \tilde{\bm{E}}_{\bm{X}_t}^{\top}\right)\right|\\
 & \qquad\qquad +\left|\mathrm{Tr}\left(\left[\left(\mathcal{A}^{*}\mathcal{A}-\mathcal{I}\right)\left( \tilde{\bm{E}}_{\bm{X}_t} \tilde{\bm{E}}_{\bm{Y}_t}^{\top}\right)\right]\tilde{\bY}_{t}\bm{\Lambda}_{t} \tilde{\bm{E}}_{\bm{X}_t}^{\top}\right)\right|\\
 &  \leq\delta_{2r}\left( \big\Vert  \tilde{\bm{E}}_{\bm{X}_t} \tilde{\bm{Y}}_{t}^{\top}\big\Vert_{\mathrm{F}}+\big\Vert \tilde{\bm{X}}_{t} \tilde{\bm{E}}_{\bm{Y}_t}^{\top}\big\Vert_{\mathrm{F}}+\big\Vert  \tilde{\bm{E}}_{\bm{X}_t} \tilde{\bm{E}}_{\bm{Y}_t}^{\top}\big\Vert_{\mathrm{F}}\right) \big\Vert \tilde{\bY}_{t}\bm{\Lambda}_{t}\tilde{\bm{E}}_{\bm{X}_t}^{\top}\big\Vert _{\mathrm{F}}.
\end{align*}
Here once again, we utilize the triangle inequality and Lemma~\ref{lemma_rip_distance}.
Noticing that $\|\bm{\Lambda}_{t}-\bm{I}\|$ is small, we further
have 
\begin{align*}
\big\Vert \tilde{\bY}_{t}\bm{\Lambda}_{t}\tilde{\bm{E}}_{\bm{X}_t}^{\top}\big\Vert _{\mathrm{F}} & \leq \big\Vert \tilde{\bY}_{t} \tilde{\bm{E}}_{\bm{X}_t}^{\top}\big\Vert _{\mathrm{F}}+\big\Vert \tilde{\bY}_{t}\left(\bm{\Lambda}_{t}-\bm{I}_{r}\right) \tilde{\bm{E}}_{\bm{X}_t}^{\top}\big\Vert _{\mathrm{F}}\\
 & \leq\big\Vert  \tilde{\bm{E}}_{\bm{X}_t} \tilde{\bm{Y}}_{t}^{\top}\big\Vert _{\mathrm{F}}+2\sigma_{\max}^{1/2}\zeta\big\Vert \tilde{\bm{E}}_{\bm{X}_t}\big\Vert _{\mathrm{F}},
\end{align*}
where we use $\|\bm{\Lambda}_{t}-\bm{I}_{r}\|\leq\zeta$ and $\|\tilde{\bm{Y}}_{t}\|\le2\sqrt{\sigma_{\max}}$.
Combine the previous two bounds and apply the basic inequality $2ab\leq a^{2}+b^{2}$
to see 
\begin{align*}
 \left|\tilde{\gamma}_{1}-\gamma_{1}\right| 
&   \leq\delta_{2r} \big\Vert \tilde{\bm{E}}_{\bm{X}_t} \tilde{\bm{Y}}_{t}^{\top}\big\Vert _{\mathrm{F}}^{2}+2\sigma_{\max}^{1/2}\zeta\delta_{2r}\big\Vert  \tilde{\bm{E}}_{\bm{X}_t} \tilde{\bm{Y}}_{t}^{\top}\big\Vert _{\mathrm{F}}\big\Vert\tilde{\bm{E}}_{\bm{X}_t}\big\Vert _{\mathrm{F}} +\delta_{2r}\big\Vert \tilde{\bm{X}}_{t}\tilde{\bm{E}}_{\bm{Y}_t}^{\top}\big\Vert _{\mathrm{F}} \big\Vert  \tilde{\bm{E}}_{\bm{X}_t} \tilde{\bm{Y}}_{t}^{\top}\big\Vert _{\mathrm{F}} \\
 &\qquad +2\sigma_{\max}^{1/2}\zeta\delta_{2r}\big\Vert \tilde{\bm{X}}_{t}\tilde{\bm{E}}_{\bm{Y}_t}^{\top}\big\Vert _{\mathrm{F}}\big\Vert\tilde{\bm{E}}_{\bm{X}_t}\big\Vert _{\mathrm{F}} +\delta_{2r} \big\Vert  \tilde{\bm{E}}_{\bm{X}_t} \tilde{\bm{E}}_{\bm{Y}_t}^{\top}\big\Vert_{\mathrm{F}}\big\Vert  \tilde{\bm{E}}_{\bm{X}_t} \tilde{\bm{Y}}_{t}^{\top}\big\Vert _{\mathrm{F}} +2\sigma_{\max}^{1/2}\zeta\delta_{2r}\big\Vert  \tilde{\bm{E}}_{\bm{X}_t} \tilde{\bm{E}}_{\bm{Y}_t}^{\top}\big\Vert _{\mathrm{F}} \big\Vert\tilde{\bm{E}}_{\bm{X}_t}\big\Vert _{\mathrm{F}}\\
 & \lesssim\delta_{2r}\left( \big\Vert  \tilde{\bm{E}}_{\bm{X}_t} \tilde{\bm{Y}}_{t}^{\top}\big\Vert _{\mathrm{F}}^{2}+\big\Vert \tilde{\bm{X}}_{t}\tilde{\bm{E}}_{\bm{Y}_t}^{\top}\big\Vert _{\mathrm{F}}^{2}+\sigma_{\min}\mathrm{dist}\left(\bm{Z}_{t},\bm{Z}_{\star}\right)\right) ,\\
 & \ll \big\Vert  \tilde{\bm{E}}_{\bm{X}_t} \tilde{\bm{Y}}_{t}^{\top}\big\Vert _{\mathrm{F}}^{2}+\big\Vert \tilde{\bm{X}}_{t}\tilde{\bm{E}}_{\bm{Y}_t}^{\top}\big\Vert _{\mathrm{F}}^{2}+\sigma_{\min}\mathrm{dist}\left(\bm{Z}_{t},\bm{Z}_{\star}\right)
\end{align*}
as long as $\delta_{2r}$ is sufficiently small. The same bound applies
to $|\tilde{\gamma}_{2}-{\gamma}_{2}|$. As a result, as long
as $\delta_{2r}$ is small enough, $\tilde{\gamma}_{1}+\tilde{\gamma}_{2}$
is lower bounded on the same order as $\gamma_{1}+{\gamma}_{2}$,
say
\begin{equation*}
\tilde{\gamma}_{1}+\tilde{\gamma}_{2}\geq\frac{1}{2} \left(\big\Vert \tilde{\bm{Y}}_{t}\tilde{\bm{E}}_{\bm{X}_t}^{\top}\big\Vert _{\mathrm{F}}^{2}+\big\Vert \tilde{\bm{X}}_{t}\tilde{\bm{E}}_{\bm{Y}_t}^{\top}\big\Vert _{\mathrm{F}}^{2}\right)   -\frac{\sigma_{\min}}{6}\mathrm{dist}^{2}\left(\bm{Z}_{t},\bm{Z}_{\star}\right).
\end{equation*}
One can then repeat the same arguments for the matrix factorization
case to obtain the linear convergence. For the sake of space, we omit
it.

%% file: conclusions.tex
\section{Conclusions}

This paper establishes the local linear convergence of gradient descent
for asymmetric low-rank matrix sensing without explicit regularization
of factor balancedness under the standard RIP assumption, as long
as a balanced initialization is provided in the basin of attraction. Coupled with the standard spectral initialization, this leads to the global convergence guarantee of the balancing-free gradient descent algorithm for asymmetric low-rank matrix sensing.
 Different from previous
work, we analyzed a new error metric that takes into account the ambiguity
due to invertible transforms, and showed that it contracts linearly
even without local restricted strong convexity. We believe that our technique
can be used for other low-rank matrix estimation problems. To conclude,
we outline a few  future research directions. 
\begin{itemize}
\item {\em Low-rank matrix completion}. We believe it is possible to
extend our analysis to study rectangular matrix completion without
regularization, by combining the leave-one-out technique in \cite{ma2017implicit,chen2019nonconvex}
to carefully bound the incoherence of the iterates for both factors
even without explicit balancing. 
\item {\em Improving dependence on $\kappa$ and $r$}. The current paper
does not try to optimize the dependence with respect to $\kappa$
and $r$ in terms of sample complexity and the size of the basin of
attraction, which are slightly worse than their regularized counterparts.
A finer analysis will likely lead to better dependencies, which we
leave to the future work.



\end{itemize}

%% file: appendix.tex
\section{Proof of Lemma~\ref{lemma:alignment-near-rotation}}

\label{proof_lemma:alignment-near-rotation} For notational convenience,
we define the following function 
\begin{equation} \label{eq:def_g}
g\left(\bm{Q}\right)\triangleq\left\Vert \bm{X}\bm{Q}-\bm{X}_{\star}\right\Vert _{\mathrm{F}}^{2}+\left\Vert \bm{Y}\bm{Q}^{-\top}-\bm{Y}_{\star}\right\Vert _{\mathrm{F}}^{2}.
\end{equation}
Clearly, the optimal alignment matrix, if exists, must be $\argmin\,g(\bm{P})$.
With this notation in place, we consider the following constrained
minimization problem: 
\begin{align*}
\min_{\bm{Q}\in\mathbb{R}^{r\times r}:\bm{Q}\text{ is invertible}} & \qquad g\left(\bm{Q}\right)\\
\text{subject to} & \qquad\left\Vert \bm{Q}-\bm{P}\right\Vert _{\mathrm{F}}\leq\frac{5\delta}{\sigma_{\min}\left(\bm{X}_{\star}\right)}.
\end{align*}
In view of Weyl's inequality, we obtain that for any feasible $\bm{Q}$,
\[
\sigma_{\min}\left(\bm{Q}\right)\geq\sigma_{\min}\left(\bm{P}\right)-\frac{5\delta}{\sigma_{\min}\left(\bm{X}_{\star}\right)}\geq\frac{1}{2}-\frac{1}{4}=\frac{1}{4}
\]
as long as $\delta\leq\sigma_{\min}(\bm{X}_{\star})/80$. As a result,
one sees that $g(\bm{Q})$ is a continuous function over $\{\bm{Q}:\|\bm{Q}-\bm{P}\|\leq5\delta/\sigma_{\min}(\bm{X}_{\star})\}$,
which is a compact set over invertible matrices. Applying the Weierstrass
extreme value theorem yields the claim that the minimizer of the constrained
problem exists. Denote this minimizer by $\bm{Q}_{1}$. In what follows,
we intend to show that $\bm{Q}_{1}$ is also the minimizer of the
unconstrained problem. Letting $\bm{Q}$ be an arbitrary matrix with
$g(\bm{Q})\leq2\delta^{2}$ (the existence is assured since $g(\bm{Q}_{1})\leq g(\bm{P})\leq2\delta^{2}$),
we have
\[
\sqrt{2}\delta\ge\left\Vert \bm{X}\bm{Q}-\bm{X}_{\star}\right\Vert _{\mathrm{F}}\ge\Vert\bm{X}\bm{Q}-\bm{X}\bm{P}\Vert_{\mathrm{F}}-\Vert\bm{X}\bm{P}-\bm{X}_{\star}\Vert_{\mathrm{F}},
\]
which in conjunction with \eqref{eq:assumption_alignment_near_rotation}
implies 
\begin{equation}
(1+\sqrt{2})\delta\geq\Vert\bm{X}(\bm{Q}-\bm{P})\Vert_{\mathrm{F}}\geq\sigma_{\min}\left(\bm{X}\right)\left\Vert \bm{Q}_{1}-\bm{P}\right\Vert _{\mathrm{F}}.\label{eq:fro-upper}
\end{equation}

We now turn to investigating $\sigma_{\min}(\bm{X})$. Weyl's inequality
tells us that 
\begin{align*}
\left|\sigma_{\min}\left(\bm{X}\bm{P}\right)-\sigma_{\min}\left(\bm{X}_{\star}\right)\right| & \leq\left\Vert \bm{X}\bm{P}-\bm{X}_{\star}\right\Vert _{\mathrm{F}}  \leq\delta\leq\frac{1}{4}\sigma_{\min}\left(\bm{X}_{\star}\right),
\end{align*}
which further implies 
\begin{align*}
\frac{3}{4}\sigma_{\min}\left(\bm{X}_{\star}\right)& \leq\sigma_{\min}\left(\bm{X}\bm{P}\right) \leq\sigma_{\min}\left(\bm{X}\right)\sigma_{\max}\left(\bm{P}\right)\leq\frac{3}{2}\sigma_{\min}\left(\bm{X}\right).
\end{align*}
Therefore we arrive at $\sigma_{\min}(\bm{X})\geq\sigma_{\min}(\bm{X}_{\star})/2$.
Putting this back to (\ref{eq:fro-upper}) yields 
which finally gives 
\[
\left\Vert \bm{Q}-\bm{P}\right\Vert _{\mathrm{F}}\le2(1+\sqrt{2})\frac{\delta}{\sigma_{\min}(\bm{X}_{\star})}<\frac{5\delta}{\sigma_{\min}(\bm{X}_{\star})}.
\]
In all, the above arguments reveal that any matrix $\bm{Q}$ such
that $g(\bm{Q})\leq2\delta^{2}$ must obey the above bound. Therefore
the minimizer of the constrained problem and that of the unconstrained
one coincide with each other. This finished the proof.

\section{Proof of Lemma~\ref{lemma:alignment} \label{proof_lemma:alignment} }

Recall the function $g\left(\bm{P}\right)$ defined in \eqref{eq:def_g} as
\begin{align}
g\left(\bm{P}\right) & =\left\Vert \bm{X}\bm{P}-\bm{X}_{\star}\right\Vert _{\mathrm{F}}^{2}+\left\Vert \bm{Y}\bm{P}^{-\top}-\bm{Y}_{\star}\right\Vert _{\mathrm{F}}^{2} \nonumber \\
 & =\mathrm{Tr}\left(\bm{X}\bm{P}\bm{P}^{\top}\bm{X}^{\top}\right)-2\mathrm{Tr}\left(\bm{P}^{\top}\bm{X}^{\top}\bm{X}_{\star}\right) +\mathrm{Tr}\left(\bm{X}_{\star}^{\top}\bm{X}_{\star}\right)+\mathrm{Tr}\left(\bm{P}^{-1}\bm{Y}^{\top}\bm{Y}\bm{P}^{-\top}\right)\nonumber \\
 & \qquad-2\mathrm{Tr}\left(\bm{P}^{-1}\bm{Y}^{\top}\bm{Y}_{\star}\right)+\mathrm{Tr}\left(\bm{Y}_{\star}^{\top}\bm{Y}_{\star}\right).\nonumber 
\end{align}
The gradient is given by 
\begin{align*}
\nabla & g\left(\bm{P}\right)  =2\bm{X}^{\top}\bm{X}\bm{P}-2\bm{X}^{\top}\bm{X}_{\star} -2\left(\bm{P}\bm{P}^{\top}\right)^{-1}\bm{Y}^{\top}\bm{Y}\left(\bm{P}\bm{P}^{\top}\right)^{-1}\bm{P}+2\bm{P}^{-\top}\bm{Y}_{\star}^{\top}\bm{Y}\bm{P}^{-\top}.
\end{align*}
Since $\bm{Q}$ minimizes $g(\bm{P})$, it must satisfy the first-order
optimality condition, i.e. 
\[
\nabla g\left(\bm{Q}\right)=\bm{0}.
\]
Identify $\tilde{\bm{X}}=\bm{X}\bm{Q}$ and $\tilde{\bm{Y}}=\bm{Y}\bm{Q}^{-\top}$
to yield the condition
\begin{align*}
\tilde{\bm{X}}^{\top}\tilde{\bm{X}}-\tilde{\bm{X}}^{\top}\bm{X}_{\star} & =\tilde{\bm{Y}}^{\top}\tilde{\bm{Y}}-\bm{Y}_{\star}^{\top}\tilde{\bm{Y}}.
\end{align*}

\section{Proof of Lemma~\ref{lemma:gradient-dominance}}

We prove the first part and the second part follows by symmetry. Denote
$\bm{E}_{x}=\bm{X}-\bm{X}_{\star}$ and $\bm{E}_{y}=\bm{Y}-\bm{Y}_{\star}$.
We have 
\[
\bm{X}\bm{Y}^{\top}-\bm{M}_{\star}=\bm{E}_{x}\bm{Y}^{\top}+\bm{X}_{\star}\bm{E}_{y}^{\top}.
\]
Since $\bm{Z}$ is aligned with $\bm{Z}_{\star}$, Lemma~\ref{lemma:alignment}
tells us that 
$$\bm{X}^{\top}\bm{E}_{x}=\bm{E}_{y}^{\top}{\bm{Y}}.$$
As a result, one has 
\begin{align}
 \left\langle \bm{X}-\bm{X}_{\star},\left(\bm{X}\bm{Y}^{\top}-\bm{M}_{\star}\right)\bm{Y}\right\rangle  
&  =\mathrm{Tr}\left(\bm{E}_{x}^{\top}\left(\bm{E}_{x}\bm{Y}^{\top}+\bm{X}_{\star}\bm{E}_{y}^{\top}\right)\bm{Y}\right)\nonumber \\
 &  =\mathrm{Tr}\left(\bm{E}_{x}^{\top}\bm{E}_{x}\bm{Y}^{\top}\bm{Y}\right)+\mathrm{Tr}\left(\bm{E}_{x}^{\top}\bm{X}_{\star}\bm{E}_{y}^{\top}\bm{Y}\right)\nonumber \\
 &  =\left\Vert \bm{Y}\bm{E}_{x}^{\top}\right\Vert _{\mathrm{F}}^{2}+\mathrm{Tr}\left(\bm{E}_{y}^{\top}\bm{Y}\bm{E}_{x}^{\top}\bm{X}\right)-\mathrm{Tr}\left(\bm{E}_{y}^{\top}\bm{Y}\bm{E}_{x}^{\top}\bm{E}_{x}\right)\nonumber \\
 &  =\left\Vert \bm{Y}\bm{E}_{x}^{\top}\right\Vert _{\mathrm{F}}^{2}+\|\bm{X}^{\top}\bm{E}_{x}\|_{\mathrm{F}}^{2}-\mathrm{Tr}\left(\bm{X}^{\top}\bm{E}_{x}\bm{E}_{x}^{\top}\bm{E}_{x}\right).\label{eq:gradient_dominance_x}
\end{align}
Complete the squares to see that 
\begin{equation*}
\|\bm{X}^{\top}\bm{E}_{x}\|_{\mathrm{F}}^{2}-\mathrm{Tr}\left(\bm{X}^{\top}\bm{E}_{x}\bm{E}_{x}^{\top}\bm{E}_{x}\right) 
 =\big\Vert \bm{E}_{x}^{\top}\bm{X}-\frac{1}{2}\bm{E}_{x}^{\top}\bm{E}_{x}\big\Vert _{\mathrm{F}}^{2}  -\frac{1}{4}\left\Vert \bm{E}_{x}\right\Vert _{\mathrm{F}}^{4}.
\end{equation*}
Combine the previous two bounds to yield the desired result. 

\section{Proof of Lemma~\ref{lemma:smoothness}}

Again, we demonstrate the claim on $\bm{X}$ and the claim on $\bm{Y}$
follows by symmetry. Given the decomposition 
\begin{align*}
\bm{X}\bm{Y}^{\top}-\bm{M}_{\star}=(\bm{X}-\bm{X}_{\star})\bm{Y}^{\top} &+\bm{X}(\bm{Y}-\bm{Y}_{\star})^{\top}  +(\bm{X}_{\star}-\bm{X})(\bm{Y}-\bm{Y}_{\star})^{\top},
\end{align*}
we obtain 
\begin{align*}
 \left\Vert \left(\bm{X}\bm{Y}^{\top}-\bm{M}_{\star}\right)\bm{Y}\right\Vert _{\mathrm{F}}  &\le\sigma_{1}(\bm{Y})\left\Vert \bm{X}\bm{Y}^{\top}-\bm{M}_{\star}\right\Vert _{\mathrm{F}}\\
 &  \le\frac{3}{2}\sigma_{1}(\bm{Y}_{\star})\Big(\left\Vert (\bm{X}-\bm{X}_{\star})\bm{Y}^{\top}\right\Vert _{\mathrm{F}} +\left\Vert \bm{X}(\bm{Y}-\bm{Y}_{\star})^{\top}\right\Vert _{\mathrm{F}}+\left\Vert (\bm{X}_{\star}-\bm{X})(\bm{Y}-\bm{Y}_{\star})^{\top}\right\Vert _{\mathrm{F}}\Big),
\end{align*}
where the last line combines the triangle inequality and Weyl's inequality
\[
\sigma_{1}\left(\bm{Y}\right)\leq\sigma_{1}(\bm{Y}_{\star})+\left\Vert \bm{Y}-\bm{Y}_{\star}\right\Vert \leq\frac{3}{2}\sigma_{1}\left(\bm{Y}_{\star}\right).
\]
The proof is then finished. 

%

%% file: RectangularMF_arxiv.bbl
\begin{thebibliography}{10}
\providecommand{\url}[1]{#1}
\csname url@samestyle\endcsname
\providecommand{\newblock}{\relax}
\providecommand{\bibinfo}[2]{#2}
\providecommand{\BIBentrySTDinterwordspacing}{\spaceskip=0pt\relax}
\providecommand{\BIBentryALTinterwordstretchfactor}{4}
\providecommand{\BIBentryALTinterwordspacing}{\spaceskip=\fontdimen2\font plus
\BIBentryALTinterwordstretchfactor\fontdimen3\font minus
  \fontdimen4\font\relax}
\providecommand{\BIBforeignlanguage}[2]{{%
\expandafter\ifx\csname l@#1\endcsname\relax
\typeout{** WARNING: IEEEtran.bst: No hyphenation pattern has been}%
\typeout{** loaded for the language `#1'. Using the pattern for}%
\typeout{** the default language instead.}%
\else
\language=\csname l@#1\endcsname
\fi
#2}}
\providecommand{\BIBdecl}{\relax}
\BIBdecl

\bibitem{CanTao10}
E.~Cand\`es and T.~Tao, ``The power of convex relaxation: Near-optimal matrix
  completion,'' \emph{IEEE Transactions on Information Theory}, vol.~56, no.~5,
  pp. 2053 --2080, May 2010.

\bibitem{chen2018harnessing}
Y.~Chen and Y.~Chi, ``Harnessing structures in big data via guaranteed low-rank
  matrix estimation: Recent theory and fast algorithms via convex and nonconvex
  optimization,'' \emph{IEEE Signal Processing Magazine}, vol.~35, no.~4, pp.
  14 -- 31, 2018.

\bibitem{davenport2016overview}
M.~A. Davenport and J.~Romberg, ``An overview of low-rank matrix recovery from
  incomplete observations,'' \emph{IEEE Journal of Selected Topics in Signal
  Processing}, vol.~10, no.~4, pp. 608--622, 2016.

\bibitem{burer2003nonlinear}
S.~Burer and R.~Monteiro, ``A nonlinear programming algorithm for solving
  semidefinite programs via low-rank factorization,'' \emph{Mathematical
  Programming}, vol.~95, no.~2, pp. 329--357, 2003.

\bibitem{bhojanapalli2016dropping}
S.~Bhojanapalli, A.~Kyrillidis, and S.~Sanghavi, ``Dropping convexity for
  faster semi-definite optimization,'' in \emph{Conference on Learning Theory},
  2016, pp. 530--582.

\bibitem{boumal2016non}
N.~Boumal, V.~Voroninski, and A.~Bandeira, ``The non-convex {B}urer-{M}onteiro
  approach works on smooth semidefinite programs,'' in \emph{Advances in Neural
  Information Processing Systems}, 2016, pp. 2757--2765.

\bibitem{tu2015low}
S.~Tu, R.~Boczar, M.~Simchowitz, M.~Soltanolkotabi, and B.~Recht, ``Low-rank
  solutions of linear matrix equations via procrustes flow,'' in
  \emph{International Conference Machine Learning}, 2016, pp. 964--973.

\bibitem{zheng2016convergence}
Q.~Zheng and J.~Lafferty, ``Convergence analysis for rectangular matrix
  completion using {B}urer-{M}onteiro factorization and gradient descent,''
  \emph{arXiv preprint arXiv:1605.07051}, 2016.

\bibitem{park2018finding}
D.~Park, A.~Kyrillidis, C.~Caramanis, and S.~Sanghavi, ``Finding low-rank
  solutions via nonconvex matrix factorization, efficiently and provably,''
  \emph{SIAM Journal on Imaging Sciences}, vol.~11, no.~4, pp. 2165--2204,
  2018.

\bibitem{chi2018low}
Y.~Chi, ``Low-rank matrix completion,'' \emph{IEEE Signal Processing Magazine},
  vol.~35, no.~5, pp. 178--181, 2018.

\bibitem{yi2016fast}
X.~Yi, D.~Park, Y.~Chen, and C.~Caramanis, ``Fast algorithms for robust {PCA}
  via gradient descent,'' in \emph{Advances in neural information processing
  systems}, 2016, pp. 4152--4160.

\bibitem{chen2020spectral}
Y.~Chen, Y.~Chi, J.~Fan, and C.~Ma, ``Spectral methods for data science: A
  statistical perspective,'' \emph{Foundations and Trends in Machine Learning},
  2020, preprint.

\bibitem{recht2010guaranteed}
B.~Recht, M.~Fazel, and P.~A. Parrilo, ``Guaranteed minimum-rank solutions of
  linear matrix equations via nuclear norm minimization,'' \emph{SIAM review},
  vol.~52, no.~3, pp. 471--501, 2010.

\bibitem{oymak2018sharp}
S.~Oymak, B.~Recht, and M.~Soltanolkotabi, ``Sharp time--data tradeoffs for
  linear inverse problems,'' \emph{IEEE Transactions on Information Theory},
  vol.~64, no.~6, pp. 4129--4158, 2018.

\bibitem{candes2009exact}
E.~J. Cand{\`e}s and B.~Recht, ``Exact matrix completion via convex
  optimization,'' \emph{Foundations of Computational Mathematics}, vol.~9,
  no.~6, pp. 717--772, 2009.

\bibitem{CanPla10}
E.~J. Cand\`es and Y.~Plan, ``Matrix completion with noise,'' \emph{Proceedings
  of the IEEE}, vol.~98, no.~6, pp. 925 --936, June 2010.

\bibitem{Negahban2012restricted}
S.~Negahban and M.~Wainwright, ``Restricted strong convexity and weighted
  matrix completion: Optimal bounds with noise,'' \emph{The Journal of Machine
  Learning Research}, vol. 98888, pp. 1665--1697, May 2012.

\bibitem{Gross2011recovering}
D.~Gross, ``Recovering low-rank matrices from few coefficients in any basis,''
  \emph{IEEE Transactions on Information Theory}, vol.~57, no.~3, pp.
  1548--1566, March 2011.

\bibitem{Recht2009SimplerMC}
B.~Recht, ``A simpler approach to matrix completion,'' \emph{Journal of Machine
  Learning Research}, vol.~12, pp. 3413--3430, Feburary 2011.

\bibitem{chen2013robustSpectralMC}
Y.~Chen and Y.~Chi, ``Robust spectral compressed sensing via structured matrix
  completion,'' \emph{IEEE Transactions on Information Theory}, vol.~60,
  no.~10, pp. 6576--6601, 2014.

\bibitem{negahban2011estimation}
S.~Negahban and M.~J. Wainwright, ``Estimation of (near) low-rank matrices with
  noise and high-dimensional scaling,'' \emph{The Annals of Statistics},
  vol.~39, no.~2, pp. 1069--1097, 2011.

\bibitem{zheng2015convergent}
Q.~Zheng and J.~Lafferty, ``A convergent gradient descent algorithm for rank
  minimization and semidefinite programming from random linear measurements,''
  in \emph{Advances in Neural Information Processing Systems}, 2015, pp.
  109--117.

\bibitem{keshavan2010few}
R.~Keshavan, A.~Montanari, and S.~Oh, ``Matrix completion from a few entries,''
  \emph{IEEE Transactions on Information Theory}, vol.~56, no.~6, pp.
  2980--2998, 2010.

\bibitem{sun2015guaranteed}
R.~Sun and Z.-Q. Luo, ``Guaranteed matrix completion via nonconvex
  factorization,'' in \emph{Symposium on Foundations of Computer Science
  (FOCS)}.\hskip 1em plus 0.5em minus 0.4em\relax IEEE, 2015, pp. 270--289.

\bibitem{jain2013low}
P.~Jain, P.~Netrapalli, and S.~Sanghavi, ``Low-rank matrix completion using
  alternating minimization,'' in \emph{Proceedings of the forty-fifth annual
  ACM symposium on Theory of computing}.\hskip 1em plus 0.5em minus 0.4em\relax
  ACM, 2013, pp. 665--674.

\bibitem{ma2017implicit}
C.~Ma, K.~Wang, Y.~Chi, and Y.~Chen, ``Implicit regularization in nonconvex
  statistical estimation: Gradient descent converges linearly for phase
  retrieval, matrix completion, and blind deconvolution,'' \emph{Foundations of
  Computational Mathematics}, pp. 1--182, 2019.

\bibitem{chen2015fast}
Y.~Chen and M.~J. Wainwright, ``Fast low-rank estimation by projected gradient
  descent: General statistical and algorithmic guarantees,'' \emph{arXiv
  preprint arXiv:1509.03025}, 2015.

\bibitem{li2018nonconvex}
Y.~Li, C.~Ma, Y.~Chen, and Y.~Chi, ``Nonconvex matrix factorization from
  rank-one measurements,'' in \emph{The 22nd International Conference on
  Artificial Intelligence and Statistics}, 2019, pp. 1496--1505.

\bibitem{li2016deconvolution}
X.~Li, S.~Ling, T.~Strohmer, and K.~Wei, ``Rapid, robust, and reliable blind
  deconvolution via nonconvex optimization,'' \emph{Applied and computational
  harmonic analysis}, vol.~47, no.~3, pp. 893--934, 2019.

\bibitem{chen2018gradient}
Y.~Chen, Y.~Chi, J.~Fan, and C.~Ma, ``Gradient descent with random
  initialization: Fast global convergence for nonconvex phase retrieval,''
  \emph{Mathematical Programming}, pp. 1--33, 2018.

\bibitem{chi2018nonconvex}
Y.~Chi, Y.~M. Lu, and Y.~Chen, ``Nonconvex optimization meets low-rank matrix
  factorization: An overview,'' \emph{IEEE Transactions on Signal Processing},
  vol.~67, no.~20, pp. 5239--5269, 2019.

\bibitem{li2017nonconvex}
Y.~Li, Y.~Chi, H.~Zhang, and Y.~Liang, ``Non-convex low-rank matrix recovery
  with arbitrary outliers via median-truncated gradient descent,''
  \emph{Information and Inference: A Journal of the IMA}, vol.~9, no.~2, pp.
  289--325, 2020.

\bibitem{zhang2018fast}
X.~Zhang, S.~Du, and Q.~Gu, ``Fast and sample efficient inductive matrix
  completion via multi-phase procrustes flow,'' in \emph{International
  Conference on Machine Learning}, 2018, pp. 5751--5760.

\bibitem{chen2019nonconvex}
J.~Chen, D.~Liu, and X.~Li, ``Nonconvex rectangular matrix completion via
  gradient descent without $\ell_{2,\infty}$ regularization,'' \emph{IEEE
  Transactions on Information Theory}, vol.~66, no.~9, pp. 5806--5841, 2020.

\bibitem{du2018algorithmic}
S.~S. Du, W.~Hu, and J.~D. Lee, ``Algorithmic regularization in learning deep
  homogeneous models: Layers are automatically balanced,'' in \emph{Advances in
  Neural Information Processing Systems}, 2018, pp. 384--395.

\bibitem{charisopoulos2019low}
V.~Charisopoulos, Y.~Chen, D.~Davis, M.~D{\'\i}az, L.~Ding, and
  D.~Drusvyatskiy, ``Low-rank matrix recovery with composite optimization: good
  conditioning and rapid convergence,'' \emph{arXiv preprint arXiv:1904.10020},
  2019.

\bibitem{ge2016matrix}
R.~Ge, J.~D. Lee, and T.~Ma, ``Matrix completion has no spurious local
  minimum,'' in \emph{Advances in Neural Information Processing Systems}, 2016,
  pp. 2973--2981.

\bibitem{ge2017no}
R.~Ge, C.~Jin, and Y.~Zheng, ``No spurious local minima in nonconvex low rank
  problems: A unified geometric analysis,'' in \emph{International Conference
  on Machine Learning}, 2017, pp. 1233--1242.

\bibitem{zhu2017global}
Z.~Zhu, Q.~Li, G.~Tang, and M.~B. Wakin, ``Global optimality in low-rank matrix
  optimization,'' \emph{IEEE Transactions on Signal Processing}, vol.~66,
  no.~13, pp. 3614--3628, 2018.

\bibitem{zhu2018global}
Z.~Zhu, D.~Soudry, Y.~C. Eldar, and M.~B. Wakin, ``The global optimization
  geometry of shallow linear neural networks,'' \emph{Journal of Mathematical
  Imaging and Vision}, pp. 1--14, 2019.

\bibitem{li2019symmetry}
X.~Li, J.~Lu, R.~Arora, J.~Haupt, H.~Liu, Z.~Wang, and T.~Zhao, ``Symmetry,
  saddle points, and global optimization landscape of nonconvex matrix
  factorization,'' \emph{IEEE Transactions on Information Theory}, vol.~65,
  no.~6, pp. 3489--3514, 2019.

\bibitem{li2018non}
Q.~Li, Z.~Zhu, and G.~Tang, ``The non-convex geometry of low-rank matrix
  optimization,'' \emph{Information and Inference: A Journal of the IMA},
  vol.~8, no.~1, pp. 51--96, 2019.

\bibitem{li2020global}
S.~{Li}, Q.~{Li}, Z.~{Zhu}, G.~{Tang}, and M.~B. {Wakin}, ``The global geometry
  of centralized and distributed low-rank matrix recovery without
  regularization,'' \emph{IEEE Signal Processing Letters}, vol.~27, pp.
  1400--1404, 2020.

\bibitem{tong2020accelerating}
T.~Tong, C.~Ma, and Y.~Chi, ``Accelerating ill-conditioned low-rank matrix
  estimation via scaled gradient descent,'' \emph{arXiv preprint
  arXiv:2005.08898}, 2020.

\bibitem{tong2020low}
------, ``Low-rank matrix recovery with scaled subgradient methods: Fast and
  robust convergence without the condition number,'' \emph{arXiv preprint
  arXiv:2010.13364}, 2020.

\bibitem{ma2019beyond}
C.~Ma, Y.~Li, and Y.~Chi, ``Beyond {P}rocrustes: Balancing-free gradient
  descent for asymmetric low-rank matrix sensing,'' in \emph{2019 53rd Asilomar
  Conference on Signals, Systems, and Computers}.\hskip 1em plus 0.5em minus
  0.4em\relax IEEE, 2019, pp. 721--725.

\end{thebibliography}
